\documentclass[12pt, draftclsnofoot, journal, letterpaper, onecolumn]{IEEEtran}

\makeatletter
\def\ps@headings{%
\def\@oddhead{\mbox{}\scriptsize\rightmark \hfil \thepage}%
\def\@evenhead{\scriptsize\thepage \hfil \leftmark\mbox{}}%
\def\@oddfoot{}%
\def\@evenfoot{}}
\makeatother 
\IEEEoverridecommandlockouts
\usepackage{bbm}
\usepackage{amsfonts}
\usepackage[dvips]{graphicx}
\usepackage{times}
\usepackage{cite}
\usepackage{amsmath}
\usepackage{array}
\usepackage{amssymb}

\newcommand{\bs}{\boldsymbol}

\usepackage{stfloats}
\usepackage{graphicx}
\usepackage{footnote}
\usepackage{booktabs}
\usepackage{array}
\usepackage{algorithm}
\usepackage{subeqnarray}
\usepackage{cases}
\usepackage{threeparttable}
\usepackage{color}
\usepackage{hyperref}
\usepackage{epstopdf}
\usepackage{algpseudocode}
\usepackage{bm}
\usepackage{multirow}
\usepackage[labelformat=simple]{subcaption}
\usepackage{adjustbox}
\usepackage{enumitem}

\usepackage{xcolor,etoolbox}

\newenvironment{answer}{%
   \color{black}
}{}

\let\mybibitem\bibitem
\renewcommand{\bibitem}[1]{%
\ifstrequal{#1}{8778671}{\color{black}\mybibitem{#1}}
{\ifstrequal{#1}{wang2018preempt}{\color{black}\mybibitem{#1}}
{\ifstrequal{#1}{kavitha2018controlling}{\color{black}\mybibitem{#1}}
{\color{black}\mybibitem{#1}}}}%
}

\newtheorem{theorem}{\bf Theorem}

\newtheorem{lemma}{\bf Lemma}
\newtheorem{definition}{\bf Definition}

\allowdisplaybreaks

\begin{document}

\title{\huge Minimum Age of Information in the Internet of Things with Non-uniform Status Packet Sizes}

\author{Bo~Zhou,~\IEEEmembership{Member,~IEEE} and Walid~Saad,~\IEEEmembership{Fellow,~IEEE}
\thanks{This work was supported by the Office of Naval Research (ONR) under MURI Grant N00014-19-1-2621 and, in part, by the National Science Foundation under Grant CNS-1836802.

% This research was supported by the U.S. Office of Naval Research (ONR) under Grant N00014-15-1-2709 and, in part, by the National Science Foundation under Grant CNS-1836802.

A preliminary version of this work  has been presented at IEEE ICC 2019\cite{AoI_ICC19}.

B.~Zhou and W.~Saad are with Wireless@VT, Bradley Department of Electrical and Computer Engineering, Virginia Tech, Blacksburg, VA 24061, USA.
Email: \{ecebo, walids\}@vt.edu}
}

\maketitle
 % \vspace{-.5cm}

\begin{abstract}
% The freshness of status information is critical for real-time Internet of Things (IoT)  applications.
In this paper, a real-time Internet of Things (IoT) monitoring  system is considered in which the IoT devices are scheduled to sample \textcolor{black}{associated} underlying physical processes and send the status updates to a common destination.
In a real-world IoT, due to the possibly different dynamics of each physical process, the sizes of the status updates for different devices are often different and each status update typically requires multiple transmission slots.
By taking into account  such multi-time slot transmissions with \emph{non-uniform sizes of the status updates} under noisy channels, the problem of joint device scheduling and status sampling is studied in order to minimize the average age of information (AoI) at the destination.
This stochastic problem is formulated as an infinite horizon average cost Markov decision process (MDP).
The monotonicity of the value function of the MDP is characterized and then used to show that the optimal scheduling and sampling policy is threshold-based with respect to the AoI at each device.
% To relieve the curse of dimensionality, a low-complexity suboptimal policy is proposed using linear approximated value functions, which is shown to possess a similar structure to the optimal policy.
To overcome the curse of dimensionality, a low-complexity suboptimal policy is proposed through a semi-randomized base policy and linear approximated value functions.
The proposed suboptimal policy is shown to exhibit a similar structure to the optimal policy, which provides a structural base for its effective performance.
A structure-aware algorithm is then developed to obtain the suboptimal policy.
The analytical results are further extended to the IoT monitoring system with \emph{random status update arrivals}, for which, the optimal scheduling and sampling policy is also shown to be threshold-based with the AoI at each device.
Simulation results illustrate the structures of the optimal policy and show a near-optimal AoI performance resulting from the proposed suboptimal solution approach.

\end{abstract}

\begin{IEEEkeywords}
Internet of things, status update, age of information, optimization, scheduling.
\end{IEEEkeywords}

\section{Introduction}
Ensuring a seamless operation of real-time Internet of Things (IoT) applications\cite{5307471,7412759,8533634,8019857} requires a timely delivery of status information collected from a variety of sensors that monitor physical processes. To characterize this timeliness of information update, the notion of \emph{age of information} (AoI) has been recently proposed\cite{6195689}. The AoI is a performance metric that can precisely quantify the timeliness of the status updates transmitted by IoT devices from the perspective of the destination.
Typically, the AoI is defined as the time elapsed since the most recently received status update was originally generated at the IoT device.
As a result, the AoI jointly accounts for the latency in sending status updates and the generation time of each status update which differentiates it from conventional performance measures, such as delay and throughput\cite{yates2016age}.

The AoI has been recently studied  under various communication system settings\cite{8000687,sun2018sampling,bedewy2018age,HsuISIT,jiang2018can,8778671,7541765,feng2018minimizing,ceran2018reinforcement,Talak,8486307}.
The authors in \cite{8000687} and \cite{sun2018sampling} propose optimal update generating policies to minimize the average AoI for a status update system with a single source node under general age penalty functions. %, for linear and non-linear age functions, respectively.
In \cite{bedewy2018age}, the authors propose an age-optimal sampling and scheduling policy for a multi-source status update system with random transmission times.
The works in \cite{HsuISIT} and \cite{jiang2018can} investigate the problem of AoI minimization for wireless networks with multiple users (IoT devices) and propose low-complexity index-based scheduling algorithms.
\textcolor{black}{The work in \cite{8778671} studies the joint design of the status sampling and updating processes to minimize the average AoI for an IoT monitoring system  under an energy constraint at each device. In particular, \cite{8778671}  proposes optimal and suboptimal polices for the cases of a single device and multiple devices, respectively.}
Different from \cite{8000687,sun2018sampling,bedewy2018age,HsuISIT,jiang2018can,8778671} where the transmission of the status update is assumed to be always successful, the works in \cite{7541765,feng2018minimizing,Talak,8486307,ceran2018reinforcement} consider that the status update may get lost during the transmission to the destination.
In particular, the authors in \cite{7541765} analyze the peak AoI in an M/M/1 queueing system with packet delivery error.
The authors in \cite{feng2018minimizing} introduce an optimal online status update policy to minimize the average AoI for an energy harvesting source with updating failures.
% The work in \cite{Talak} proposes optimal scheduling algorithms to minimize the average and peak AoI for wireless networks with noisy channels.
% In \cite{ceran2018reinforcement}, the authors propose an online scheduling algorithm to minimize the average AoI for a multi-user status update system with noisy channels.
% The work in \cite{Talak} proposes optimal scheduling algorithms to minimize the average and peak AoI for wireless networks with noisy channels.
The work in \cite{ceran2018reinforcement} proposes an online scheduling algorithm to minimize the average AoI for a multi-user status update system with noisy channels.
The works in \cite{Talak} and \cite{8486307} propose  optimal and low-complexity suboptimal scheduling algorithms to minimize the AoI for wireless networks with noisy channels.
The authors in \cite{feng2019age} consider  the optimal transmission scheduling to minimize the average AoI in an erasure channel with rateless codes.  
\textcolor{black}{In \cite{kavitha2018controlling}, the authors study the optimal packet drop policies that can minimize the average AoI for single-source and multiple-source information updating systems with random transmission times.}  

These existing works, e.g., \cite{8000687,sun2018sampling,bedewy2018age,HsuISIT,jiang2018can,8778671,7541765,feng2018minimizing,Talak,8486307,ceran2018reinforcement}, assume that the delivery of one status update can be done within one transmission slot and it takes the same time for different IoT devices to send their status updates to the destination.
\textcolor{black}{However, due to the limited transmission capabilities of low-power IoT devices and the rich information contained in one status update for sophisticated IoT processes, such as artificial intelligence tasks\cite{teerapittayanon2017distributed,chen2017machine}, a single status update from each IoT device may be composed of \emph{multiple transmission packets.}}
%and the rich information contained in one status update for sophisticated IoT processes, such as artificial intelligence tasks\cite{teerapittayanon2017distributed,chen2017machine}
% the delivery of a single status update from each IoT device  may take \emph{multiple transmission slots}.
Moreover, for heterogeneous IoT tasks and varying underlying processes, the sizes of the status updates collected by different  devices are often different\cite{WU201714430}.
In presence of \emph{non-uniform status update packet sizes}, a key question for each  device is whether to continue sending its current in-transmission status update or sample the underlying process and send a newly generated status update.
Prior results \cite{8000687,sun2018sampling,bedewy2018age,HsuISIT,jiang2018can,8778671,7541765,feng2018minimizing,Talak,8486307,ceran2018reinforcement}  are no longer applicable for such a scenario, as they assume uniform status update packet sizes and a network in which one status update can be delivered in one transmission slot.
\textcolor{black}{%The most relevant studies to this work is \cite{8778671} and \cite{wang2018preempt}.
Recently, the authors in \cite{wang2018preempt} proposed  optimal status update policies to minimize the average AoI for a status monitoring system with uniform and non-uniform packet sizes. 
However, the focus of \cite{wang2018preempt} is restricted to a system with a single source and random arrivals of status updates. Indeed, scenarios in which there exists multiple sources whose status updates can be generated at will by the devices, are not considered in \cite{wang2018preempt}.}
Clearly, how to minimize the AoI by enabling \textcolor{black}{multiple} IoT devices to intelligently schedule and update their status information over multiple slots per update, under non-uniform status update packet sizes, remains an open problem.
% In presence of the non-uniform status update packet sizes, it remains unknown how to intelligently schedule the multiple IoT devices to sample the physical processes and send the status updates to the destination, so as to minimize the average AoI.
% In presence of the non-uniform status update packet sizes, prior results \cite{8000687,HsuISIT,jiang2018can,8778671,7541765,feng2018minimizing,Talak,ceran2018reinforcement} are no longer applicable,
% and thus, it is still an open issue on the joint scheduling of the multiple IoT devices and sampling for each , so as to minimize the average AoI.

The main contribution of this paper is, thus, a joint design \textcolor{black}{of} the device scheduling and status sampling policy that minimizes the average AoI for a real-time IoT monitoring system with multiple IoT devices, by taking into account non-uniform sizes of status update packets under noisy channels.
\textcolor{black}{In the considered model, different IoT devices are associated with different underlying physical processes. Moreover, for each IoT device, we introduce the two concepts of AoI at the  device and  AoI at the receiver (destination) so as to measure the age of the current in-transmission update at the device and the most recently received update at the destination, respectively.}
We formulate the stochastic control problem related to the IoT device scheduling problem as an infinite horizon average cost Markov decision process (MDP).
By exploiting the special properties of the AoI dynamics, we characterize the monotonicity property of the value function for the MDP.
Then, we show that the optimal scheduling and sampling policy is threshold-based with respect to the AoI at each IoT device.
To reduce the computational complexity, we propose a low-complexity suboptimal policy, which is shown to possess a similar structure to the optimal policy.
This is achieved through a linear approximation of the value functions and a semi-randomized base policy, which can maintain the monotonicity of the value function.
Then, we propose a structure-aware algorithm to obtain the proposed policy.
Moreover, we  extend the above analytical results for the IoT system in which the status information updates randomly arrive at each IoT device, and show that the optimal scheduling and sampling policy is also threshold-based with  the AoI at each IoT device.\textcolor{black}{\footnote{\textcolor{black}{The considered two scenarios, one in which status updates are generated at will by each device, and another in which they arrive randomly at each device, are similar to the active and buffered sources (devices) considered in \cite{Talak}. However, we note that the work \cite{Talak} did not consider multiple transmission packets for a single status update.}}}
Simulation results show that, for the IoT system without random status update arrivals, the optimal policy is not threshold-based with respect to \textcolor{black}{the AoI at the receiver}, and the proposed suboptimal policy achieves a near-optimal performance and significantly outperforms the semi-randomized base policy; and for the IoT system with random status update arrivals, the device is more  willing to start transmitting the status update in the buffer when the arrival rate of the status updates is larger.

The rest of this paper is organized as follows. In Section II, we introduce the system model and the problem formulation. Section III characterizes the structural property of the optimal policy and Section IV presents a low-complexity structure-aware suboptimal solution. Section V  extends the analysis to the system with random status update arrivals and characterizes the structural property of its optimal policy.  Simulation results and analysis are provided in Section VI.
Finally, conclusions are drawn in Section VII.

\section{System Model and Problem Formulation}\label{sec:systemmodel}

\begin{figure}[!t]
\begin{centering}
\includegraphics[scale=.45]{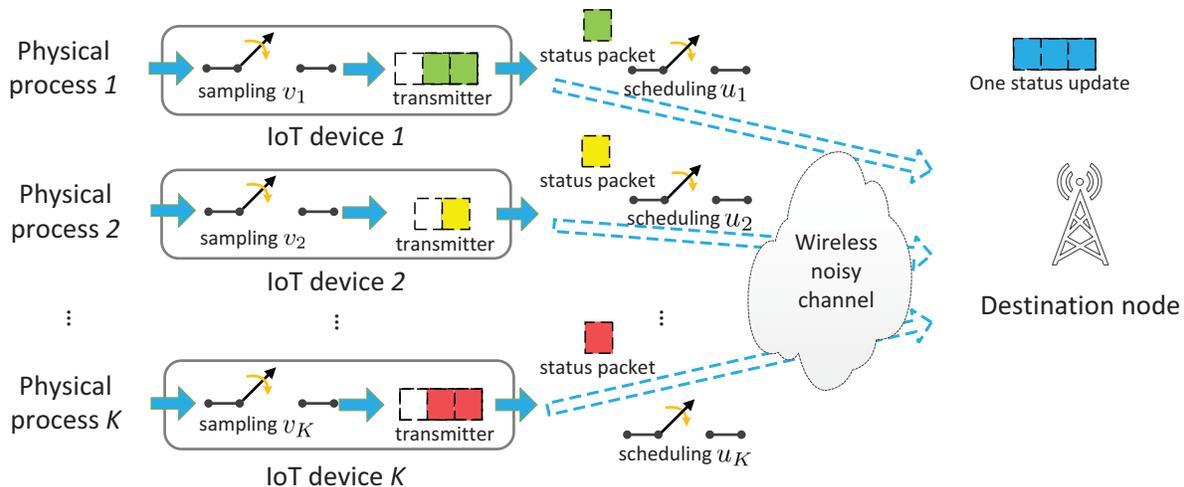}
 \caption{Illustration of a real-time IoT monitoring system with non-uniform status packet sizes and wireless noisy channels. Each status update is assumed to use more than one transmission slot to be sent to the destination.}\label{fig:system}
\end{centering}
 % \vspace{-0.35cm}
\end{figure}

Consider a real-time IoT monitoring system consisting of a set $\mathcal{K}$ of $K$ IoT devices and
 a remote destination node (e.g., a control center or base station), as illustrated in Fig.~\ref{fig:system}. The IoT devices can collect the real-time status information of the associated underlying physical processes and update the status information packets to the common destination.
  We assume that the time needed for generating the status packets is negligible for each IoT device
  \textcolor{black}{\footnote{\textcolor{black}{In Appendix~\ref{app:non-zero generation time}, we extend the considered framework to the case in which the generation time of status packets is non-zero.}}}, as done in \cite{feng2018minimizing,Talak,ceran2018reinforcement,8570843}. \textcolor{black}{This is relevant to several practical IoT applications such as environmental monitoring or surveillance with smart camera systems\cite{6195561,Chen:2013:LCS:2422966.2422978}, in which, the devices can instantaneously capture an image or a short video.}
In our model, the \emph{sizes of the status updates} for different devices can be \emph{different}, and for each device,
\textcolor{black}{one status update may be composed of \emph{several transmission packets}.}
 % \emph{several transmission slots} may be needed to transmit one status update.
 This is different from prior works\cite{8000687,sun2018sampling,bedewy2018age,HsuISIT,jiang2018can,8778671,7541765,feng2018minimizing,Talak,8486307,ceran2018reinforcement} in which the status updates of different devices are assumed to be of the same size, and for each device, one status update can be transmitted to the destination through only one transmission slot.

We consider a discrete-time system, in which time is partitioned into scheduling slots with unit duration indexed by $t=1,2,\cdots$. For each IoT device $k\in\mathcal{K}$, let $L_k\geq 2$
be the number of packets pertaining to one status update. % Without loss of generality,
We assume that each device can transmit at most one packet in one slot.
We consider that the channel between each IoT device $k$ and the destination is noisy\cite{7541765,ceran2018reinforcement,feng2018minimizing,Talak,8486307}.
Hence, the probability with which a packet sent by device $k$ is successfully delivered to the destination will be $\lambda_k\in(0,1]$, which constitutes the channel reliability for the transmission of device $k$.
% The packet transmitted during each slot will be successfully delivered to the destination with probability $\lambda_k\in(0,1]$.
As considered in\cite{7541765,ceran2018reinforcement,feng2018minimizing}, we assume that there is a perfect feedback channel between each device and the destination, such that each device will be immediately informed on whether its transmission is successful.
\subsection{Monitoring Model}
In each slot, the network has to determine which IoT devices must be scheduled so as to update their status.
For each scheduled device, because of the possible failure of each transmission and the need for multiple packets for a single status update, its current in-transmission status update may become outdated at the destination.
Thus, the network must decide whether a scheduled device continues its current in-transmission update or samples and sends a new status update.

% Then, for each scheduled device, the network must decide whether to continue the current in-transmission update or sample and transmit a new status update.
% This is due to the possible failure of each transmission and that any status update may contain multiple packets.
For each device $k$, let $u_k(t)\in\{0,1\}$ be the scheduling action at time slot $t$, where $u_{k}(t)=1$ indicates that device $k$ is scheduled to transmit its status update at slot $t$, and $u_k(t)=0$, otherwise. In each slot, we consider that at most $M\leq K$ IoT devices can update their status packets concurrently without collisions over different orthogonal channels \cite{7962670}. Mathematically, we must have $\sum_{k\in\mathcal{K}}u_k(t)$$\leq M$ for all $t$. Let $\bs{u}(t)\triangleq (u_k(t))_{k\in\mathcal{K}}\in\mathcal{U}$ be the system scheduling action at slot $t$, where $\mathcal{U}\triangleq \{(u_k)_{k\in\mathcal{K}} | u_k\in\{0,1\}~\forall k\in\mathcal{K}\text{~and~} \sum_{k\in\mathcal{K}} u_k \leq M \}$ is the feasible system scheduling action space.
 % Due to the possible failure of each transmission and that each status update may contain multiple packets, the scheduled IoT devices need to determine whether to continue the current in-transmission update or sample and transmit a new status update.
 Let $v_k(t)\in\{1,2\}$ be the sampling action for device $k$ at slot $t$, where $v_k(t)=1$ indicates that device $k$ will continue transmitting its current in-transmission update at slot $t$, and $v_k(t)=2$ indicates that device $k$ will drop the current in-transmission update and start transmitting a newly generated status update at slot $t$. For notational convenience, we set $v_k(t)=0$ if device $k$ is not scheduled at slot $t$.
Let $\bs{v}(t)\triangleq (v_k(t))_{k\in\mathcal{K}}\in\mathcal{V}\triangleq\{0,1,2\}^K$ be the system sampling action at slot $t$, where $\mathcal{V}$ is the system sampling action space.
Let $\bs{w}_k(t)\triangleq (u_k(t),v_k(t))$ be the control action vector of device $k$ at slot $t$.
\textcolor{black}{Note that, for each device, there are only three valid actions $(0,0)$, $(1,1)$, and $(1,2)$.}
Let $\bs{w}(t)\triangleq (\bs{u}(t),\bs{v}(t))\in\mathcal{W}\triangleq\mathcal{U}\times\mathcal{V}$ be the system control action at slot $t$, where $\mathcal{W}$ is the feasible system action space.
% \par\nobreak \vspace{-12pt}
\subsection{Age of Information Model}\label{sec:aoi model}

We use the AoI as the key performance metric to characterize the timeliness of the status information updates, which is defined as the time elapsed since  the most recently received update was generated.
For each device $k$, we define $A_{r,k}(t)$  \textcolor{black}{as the AoI at the receiver (destination)}  for device $k$ at the beginning of slot $t$. Assuming that the most recent update at the destination at time $t$ was generated at time $\delta_k(t)$ from device $k$, then we have $A_{r,k}(t)= t - \delta_k(t)$.
Note that \textcolor{black}{the AoI at the receiver} depends on the AoI at each device, i.e., the age of the status update of each device.
For each device $k$, we denote by $A_{d,k}(t)$ the AoI at device $k$ at the beginning of slot $t$.
Let $\hat{A}_{d,k}$ and $\hat{A}_{r,k}$ be, respectively, the upper limits of the AoI at device $k$ and the AoI for device $k$ at the destination.
For tractability\textcolor{black}{\cite[Chapter 5.6]{bertsekas4}}, we assume that $\hat{A}_{d,k}$ and $\hat{A}_{r,k}$ are finite, but can be arbitrarily large.
Let $\mathcal{A}_{d,k}\triangleq\{0,1,\cdots,\hat{A}_{d,k}\}$ and $\mathcal{A}_{r,k}\triangleq\{0,1,\cdots,\hat{A}_{r,k}\}$ be, respectively, the state space for the AoI at device $k$ and \textcolor{black}{the AoI at the receiver} for device $k$.
Since any given transmission may fail and any status update may contain multiple packets, we need to record the number of packets $D_k(t)\in\mathcal{D}_k\triangleq \{1,\cdots, L_k\}$ that  are left to be transmitted to complete the current in-transmission status update for each device $k$ at slot $t$.  Let $\bs{X}_k(t) \triangleq (A_{d,k}(t),A_{r,k}(t),D_k(t))\in\mathcal{X}_k\triangleq \mathcal{A}_{d,k}\times \mathcal{A}_{r,k}\times \mathcal{D}_{k}$ be the system state vector of device $k$ at slot $t$, where $\mathcal{X}_k$ denotes the system state space of device $k$.
Let $\bs{X}(t)\triangleq (\bs{X}_k(t))_{k\in\mathcal{K}}\in\mathcal{X}\triangleq \prod_{k\in\mathcal{K}}\mathcal{X}_k$ be the system state matrix at slot $t$, where $\mathcal{X}$ denotes the system state space.

% to save more space
% \begin{equation}\label{eqn:threshold}
%   \alpha_r^*(Q)=\left\{
%     \begin{array}{ll}
%       1, & \hbox{if $Q>Q_{th}^*$;} \\
%       0, & \hbox{otherwise.}
%     \end{array}
%   \right.
% \end{equation}

% \begin{align}\label{eqn:aoi_bs_k}
% &A_{r,k}(t+1)\\
% &=\left\{
%     \begin{array}{ll}
%  &\min\{A_{d,k}(t)+1,\hat{A}_{r,k}\},  ~\text{if}~\bs{w}_k(t)=(1,1),D_k(t)=1,\\
% &\hspace{33mm}\text{and~transmission succeeds at}~t,\\
%                 &\min\{A_{r,k}(t)+1,\hat{A}_{r,k}\}, ~ \text{otherwise.}
%     \end{array}
%   \right.
%   \nonumber
% \end{align}

When device $k$ is scheduled to continue with the current in-transmission status update at slot $t$ (i.e., $\bs{w}_k(t)=(1,1)$) and the transmission is successful, then, if there is only one remaining packet at $t$ (i.e., $D_k(t)=1)$, the number of remaining status packets will be reset to $L_k$; otherwise, the number will decrease by one. When device $k$ is scheduled to sample and transmit a new status update at slot $t$ (i.e., $\bs{w}_k(t)=(1,2)$), then the number of the remaining packets will be $L_k-1$ if the transmission is successful, and $L_k$, otherwise. Thus, for each device $k$, we can write the dynamics of $D_k(t)$:
% \begin{align}\label{eqn:d_device_k}
% &D_k(t+1)\\
% &=\begin{cases} &\mathbbm{1}(D_k(t)=1)L_k+\mathbbm{1}(D_k(t)>1)(D_k(t)-1),  \\
% &\hspace{12mm}~\text{if}~\bs{w}_k(t)=(1,1)~\text{and~transmission succeeds at}~t,\\
% &L_k-1, ~ \text{if}~ \bs{w}_k(t)=(1,2)~\text{and~transmission succeeds at}~t,\\
% % &L_k-1, ~ \text{if}~ \bs{w}_k(t)\text{=}(1,2)~\text{and~transmission succeeds at}~t\\
% &L_k, ~ \text{if}~\bs{w}_k(t)=(1,2)~\text{and~transmission fails at}~t,\\
% &D_k(t), ~ \text{otherwise.}
%       \end{cases}\nonumber
% \end{align}
\begin{align}\label{eqn:d_device_k}
D_k(t+1)
&=\begin{cases} &\mathbbm{1}(D_k(t)=1)L_k+\mathbbm{1}(D_k(t)>1)(D_k(t)-1),  \\
&\hspace{14mm}~\text{if}~\bs{w}_k(t)=(1,1)~\text{and~transmission succeeds at}~t,\\
% &L_k-1, ~ \text{if}~ \bs{w}_k(t)=(1,2)~\text{and~transmission succeeds}\\
% &\hspace{70mm}\text{ at}~t,\\
&L_k-1, ~ \text{if}~ \bs{w}_k(t)=(1,2)~\text{and~transmission succeeds at}~t,\\
&L_k, \hspace{7mm}~ \text{if}~\bs{w}_k(t)=(1,2)~\text{and~transmission fails at}~t,\\
&D_k(t), \hspace{1.2mm}~ \text{otherwise.}
      \end{cases}
\end{align}

In terms of the AoI at device $k$, when device $k$ is scheduled to continue sending its current in-transmission update at slot $t$ (i.e., $\bs{w}_k(t)=(1,1)$), if there remains only one packet and the transmission is successful, then the AoI will decrease to zero. When device $k$ is scheduled to transmit a new status update at $t$ (i.e., $\bs{w}_k(t)=(1,2)$), if the transmission fails, then the AoI will decrease to zero, otherwise, the AoI will be one.
For all remaining cases, the AoI will increase by one. Thus, the AoI dynamics of device $k$ are given by:
\begin{align}\label{eqn:aoi_device_k}
A_{d,k}(t+1)
&=\begin{cases} &0,  ~\text{if}~\bs{w}_k(t)=(1,1), D_k(t) = 1, \text{and~transmission succeeds at}~t;\\
&\hspace{8mm}~\text{or}~\bs{w}_k(t)=(1,2)~\text{and~transmission fails at}~t,\\
&1, ~ \text{if}~ \bs{w}_k(t)=(1,2)~\text{and~transmission succeeds at}~t,\\
                &\min\{A_{d,k}(t)+1,\hat{A}_{d,k}\}, ~ \text{otherwise.}
      \end{cases}
\end{align}

% \begin{align}\label{eqn:aoi_device_k}
% &A_{d,k}(t+1)\\
% &=\begin{cases} &0,  ~\text{if}~\bs{w}_k(t)=(1,1), D_k(t) = 1, \text{and~transmission}\\
% &\hspace{4mm}\text{succeeds at}~t;~\text{or}~\bs{w}_k(t)=(1,2)~\text{and~transmission}\\
% &\hspace{62mm}\text{fails at}~t,\\
% &1, ~ \text{if}~ \bs{w}_k(t)=(1,2)~\text{and~transmission succeeds at}~t,\\
%                 &\min\{A_{d,k}(t)+1,\hat{A}_{d,k}\}, ~ \text{otherwise.}
%       \end{cases}\nonumber
% \end{align}

% single column version

% \begin{align}\label{eqn:aoi_bs_k}
% &A_{r,k}(t+1)\\
% &=\begin{cases} &\min\{A_{d,k}(t)+1,\hat{A}_{r,k}\},  ~\text{if}~\bs{w}_k(t)=(1,1), D_k(t)=1,~\text{and~transmission succeeds at}~t\\
%                 &\min\{A_{r,k}(t)+1,\hat{A}_{r,k}\}, ~ \text{otherwise.}
%       \end{cases}.\nonumber
% \end{align}

For \textcolor{black}{the AoI at the receiver} of device $k$, when device $k$ is scheduled to continue sending its current in-transmission status update and only one packet remains, then the destination AoI decreases to the AoI at device $k$ at slot $t$, otherwise, it increases by one. Thus,  the dynamics of the destination's AoI for device $k$ are given by:
\begin{align}\label{eqn:aoi_bs_k}
A_{r,k}(t+1)
&=\begin{cases} &\min\{A_{d,k}(t)+1,\hat{A}_{r,k}\},  ~\text{if}~\bs{w}_k(t)=(1,1),D_k(t)=1,\text{and}\\
&\hspace{43mm}\text{transmission succeeds at}~t,\\
                &\min\{A_{r,k}(t)+1,\hat{A}_{r,k}\}, ~ \text{otherwise.}
      \end{cases}
\end{align}

Note that the dynamics in \eqref{eqn:d_device_k}-\eqref{eqn:aoi_bs_k} are highly different from the AoI dynamics in \cite{8778671}.

% \normalsize
% \begin{align}\label{eqn:aoi_bs_k}
% &A_{r,k}(t+1)\\
% &=\begin{cases} &\min\{A_{d,k}(t)+1,\hat{A}_{r,k}\},  ~\text{if}~\bs{w}_k(t)=(1,1),D_k(t)=1,\\
% &\hspace{33mm}\text{and~transmission succeeds at}~t,\\
%                 &\min\{A_{r,k}(t)+1,\hat{A}_{r,k}\}, ~ \text{otherwise.}
%       \end{cases}\nonumber
% \end{align}

\subsection{Problem Formulation}\label{sec:problem formulation}
Our goal is to study how to jointly control the IoT device scheduling and status sampling processes so as to minimize the \emph{average AoI} at the destination under non-uniform status update packet sizes and noisy channels.
Given an observed system state $\bs{X}$, the system scheduling and sampling action $\bs{w}$ is determined according to the following  policy.

\begin{definition}\label{definition:stationary_policy_IoT}
A \emph{feasible stationary scheduling and sampling policy} $\pi=(\pi_{u},\pi_{v})$ is defined as a mapping from the system state $\bs{X}\in\mathcal{X}$ to the feasible system control action $\bs{w}\in\mathcal{W}$, where $\pi_{u}(\bs{X})=\bs{u}$ and $\pi_{v}(\bs{X})=\bs{v}$.
\end{definition}

By the dynamics in \eqref{eqn:d_device_k}-\eqref{eqn:aoi_bs_k}, the induced random process $\{\bs{X}(t)\}$ for a given feasible stationary policy $\pi$ is a controlled Markov chain having the following transition probability:
\begin{equation}
\Pr[\bs{X}'|\bs{X},\bs{w}] = \prod_{k=1}^K \Pr[\bs{X}'_k|\bs{X}_k,\bs{w}_k],\label{eqn:trans_prob}
\end{equation}
where
\begin{align}\label{eqn:trans_prob_k}
&\Pr[\bs{X}'_k|\bs{X}_k,\bs{w}_k]\nonumber\\
&=\Pr[\bs{X}_k(t+1)=\bs{X}'_k|\textcolor{black}{\bs{X}_k(t)=\bs{X}_k,\bs{w}_k(t)=\bs{w}_k}]\nonumber\\
&=\begin{cases}
\lambda_k,  &~\text{if}~\bs{X}'_k = \bs{X}_{k,s}~\text{and~} u_k=1,\\
1-\lambda_k, &~\text{if}~\bs{X}'_k = \bs{X}_{k,f}~\text{and~} u_k=1,\\
 1,&~\text{if}~\bs{X}'_k = \bs{X}_{k,un}~\text{and~} u_k=0,\\
                0, &~\text{otherwise.}
      \end{cases}
\end{align}
Here, $\bs{X}_{k,s}$ and $\bs{X}_{k,f}$ indicate whether a transmission succeeds or fails, \textcolor{black}{and $\bs{X}_{k,un}$ denotes the next system state for the case in which user $k$ is not scheduled.}
According to \eqref{eqn:d_device_k}-\eqref{eqn:aoi_bs_k}, we know that, if $v_k=1$, i.e., device $k$ is scheduled to continue sending its current in-transmission update, then
\begin{align}
&\bs{X}_{k,s}
=\begin{cases}
(0,\min\{A_{d,k}+1,\hat{A}_{r,k}\}, L_k),  &\text{if}~D_k=1,\\
(\min\{A_{d,k}+1,\hat{A}_{d,k}\},\min\{A_{r,k}+1,\hat{A}_{r,k}\}, D_k-1),
&\text{otherwise.}
      \end{cases}\\
 &\bs{X}_{k,f}= (\min\{A_{d,k}+1,\hat{A}_{d,k}\},\min\{A_{r,k}+1,\hat{A}_{r,k}\}, D_k).
\end{align}
% \begin{equation}
% \bs{X}_{k,f}= (\min\{A_{d,k}+1,\hat{A}_{d,k}\},\min\{A_{r,k}+1,\hat{A}_{r,k}\}, D_k).
% \end{equation}
if $v_k=2$, i.e., device $k$ is scheduled to start sending a new status update, then
\begin{align}
&\bs{X}_{k,s}=(1,\min\{A_{r,k}+1,\hat{A}_{r,k}\}, L_k-1),\\
&\bs{X}_{k,f}= (0,\min\{A_{r,k}+1,\hat{A}_{r,k}\}, L_k),
\end{align}
 and
  %$\bs{X}_{k,un}= (\min\{A_{d,k}+1,\hat{A}_{d,k}\},\min\{A_{r,k}+1,\hat{A}_{r,k}\}, D_k)$.
\begin{align}
\bs{X}_{k,un}= (\min\{A_{d,k}+1,\hat{A}_{d,k}\},\min\{A_{r,k}+1,\hat{A}_{r,k}\}, D_k).
\end{align}

As a result, under a feasible stationary policy $\pi$, \textcolor{black}{the average AoI} at the receiver starting from a given initial state $\bs{X}(1)=\bs{X}_1\in\mathcal{X}$ is given by:
\begin{equation}
\bar{A}_r^{\pi}(\bs{X}_1)\triangleq\limsup_{T\to\infty}\frac{1}{T}\sum_{t=1}^T \sum_{k=1}^K \mathbb{E} \left[A_{r,k}(t)|\bs{X}_1\right],\label{eqn:avg_aoi}
\end{equation}
where the expectation is taken with respect to the measure induced by policy $\pi$. Note that, the analytical framework and results for the linear age function in \eqref{eqn:avg_aoi} also hold for non-decreasing non-linear age functions (see examples in \cite{sun2018sampling}).

We seek to find the optimal scheduling and sampling policy that minimizes the \emph{average AoI} at the receiver, as follows:\textcolor{black}{\footnote{\textcolor{black}{In this work, we do not explicitly consider the energy limitations on the IoT devices, due to the development of energy harvesting and battery storage technologies\cite{6951347,6158609}. However, it is possible to extend the analytical framework   to the scenario in which there are energy constraints for the IoT devices, by following a constrained MDP approach in our previous work \cite{8778671}.}}}
\begin{align}
\bar{A}_r^*(\bs{X}_1)\triangleq\textcolor{black}{\inf_{\pi}}\bar{A}_r^{\pi}(\bs{X}_1),\label{eqn:mdp}
\end{align}
where $\pi$ is a feasible stationary policy in Definition~\ref{definition:stationary_policy_IoT} and $\bar{A}_r^*(\bs{X}_1)$ denotes the \textcolor{black}{infimum} average \textcolor{black}{the AoI at the receiver starting}  from a given initial state $\bs{X}(1)=\bs{X}_1$ achieved by the optimal policy $\pi^*$.
The problem in \eqref{eqn:mdp} is an infinite horizon average cost MDP, which is challenging to solve due to the curse of dimensionality \cite{bertsekas4}. Hereinafter, as is commonly used in the literature (e.g., \cite{ceran2018reinforcement} and \cite{djonin2007mimo}), we restrict our attention to \emph{stationary unichain policies} to ensure that the optimal stationary policy exists.

\section{Structural Properties of the Optimal Policy}\label{sec:optimal policy}

According to \textcolor{black}{\cite[Propositions 5.2.1, 5.2.3, and 5.2.5]{bertsekas4}}\textcolor{black}{\footnote{\textcolor{black}{The upper limits of the AoI at the device $\hat{A}_{d,k}$ and the AoI at the receiver $\hat{A}_{r,k}$  guarantee the system state space to be finite, based on which these results in \cite{bertsekas4} can be used to prove Lemma~\ref{lemma:bellman}.}}}, the optimal scheduling and sampling policy $\pi^*$ can be obtained by solving the following Bellman equation.

\begin{lemma}\label{lemma:bellman}
 There \textcolor{black}{exists} a \textcolor{black}{unique} scalar $\theta$ and a value function $\{V(\bs{X})\})$ satisfying:
\begin{align}
  \theta + V(\bs{X}) = \sum_{k=1}^KA_{r,k} + \min_{\bs{w}\in\mathcal{W}} \sum_{\bs{X}'\in\mathcal{X}}\Pr[\bs{X}'|\bs{X},\bs{w}] V(\bs{X}'),~\forall \bs{X}\in\mathcal{X},\label{eqn:bellman}
\end{align}
where $\Pr[\bs{X}'|\bs{X},\bs{w}]$ is given by \eqref{eqn:trans_prob}.
Here, $\theta=\bar{A}_r^*$ is the optimal value to \eqref{eqn:mdp} for all initial state $\bs{X}_1\in\mathcal{X}$ and the optimal policy achieving the optimal value $\theta$ will be
\begin{equation}
  \pi^*(\bs{X})=\arg\min_{\bs{w}\in\mathcal{W}}\sum_{\bs{X}'\in\mathcal{X}}\Pr[\bs{X}'|\bs{X},\bs{w}] V(\bs{X}'),~\forall \bs{X}\in\mathcal{X}.\label{eqn:optimal_pi}
\end{equation}
\end{lemma}

From Lemma~\ref{lemma:bellman}, we can see that the optimal policy $\pi^*$ relies upon the value function $V(\cdot)$. To obtain $V(\cdot)$, we need to solve the Bellman equation in \eqref{eqn:bellman}, for which there is no closed-form solution in general. Moreover, numerical solutions such as value iteration and policy iteration do not typically provide many design insights and are usually of high  complexity due to the curse of dimensionality. Therefore, we need to study the structural properties of the optimal policy and design new structure-aware low-complexity solutions.

% Now, we characterize the structural properties of the optimal policy for the MDP in \eqref{eqn:mdp}.
First, by the dynamics in \eqref{eqn:d_device_k}-\eqref{eqn:aoi_bs_k} and using the relative value iteration algorithm, we can show the following property of the value function $V(\bs{X})$. %\footnote{All proofs are omitted due to space limitations and can be found in \cite{dropbox}.}
 Define  $\bs{A}_d\triangleq (A_{d,k})_{k\in\mathcal{K}}$, $\bs{A}_r\triangleq (A_{r,k})_{k\in\mathcal{K}}$, and $\bs{D}\triangleq (D_{k})_{k\in\mathcal{K}}$

\begin{lemma}\label{lemma:propertyV}
For any $\bs{X}^1,\bs{X}^2 \in\mathcal{X}$ such that $\bs{A}_d^2\succeq\bs{A}_d^1$, $\bs{A}_r^2\succeq\bs{A}_r^1$, and $\bs{D}^2=\bs{D}^1$, we have $V(\bs{X}^2)\geq V(\bs{X}^1)$.\footnote{The notation $\succeq$ indicates component-wise $\geq$.}
\end{lemma}

\begin{IEEEproof}
  See Appendix~\ref{app:propertyV}.
  % See Appendix~A.
\end{IEEEproof}

Then, we introduce the state-action cost function according to the right-hand side of the Bellman equation in \eqref{eqn:bellman}:
\begin{align}
  J(\bs{X},\bs{w})=\sum_{k=1}^KA_{r,k} + \sum_{\bs{X}'\in\mathcal{X}}\Pr[\bs{X}'|\bs{X},\bs{w}] V(\bs{X}').\label{eqn:J_function}
\end{align}
Based on $J(\bs{X},\bs{w})$, we further define:
\begin{align}\label{eqn:threshold}
\phi_{\bs{w}}(\bs{X}_{-d,-k})\triangleq\begin{cases}\min\Phi_{\bs{w}}(\bs{X}_{-d,-k}),  & \text{if}~\Phi_{\bs{w}}(\bs{X}_{-d,-k})\neq\emptyset, \\
            +\infty,  &\text{otherwise},
  \end{cases},
\end{align}
where $\bs{X}_{-d,-k}\triangleq \bs{X}\setminus \{A_{d,k}\}$	and $\Phi_{\bs{w}}(\bs{X}_{-d,-k}) \triangleq \{A_{d,k}|A_{d,k}\in\mathcal{A}_{d,k}\text{~and~} J(A_{d,k},\bs{X}_{-d,-k},\bs{w})\leq J(A_{d,k},\bs{X}_{-d,-k},\bs{w}')~\forall \bs{w}'\in\mathcal{W}\text{~and~}\bs{w}'\neq \bs{w}\}.$
% $\Phi_{\bs{w}}(\bs{X}_{-d,-k}) \triangleq \{A_{d,k}\in\mathcal{A}_{d,k}|J(A_{d,k},\bs{X}_{-d,-k},\bs{w})\leq J(A_{d,k},\bs{X}_{-d,-k},\bs{w}')~\forall \bs{w}'\in\mathcal{W}\text{~and~}\bs{w}'\neq \bs{w}\}$.
% \begin{align}\label{eqn:defi_phi}
% &\Phi_{\bs{w}}(\bs{X}_{-d,-k}) \triangleq \Big\{A_{d,k}|A_{d,k}\in\mathcal{A}_{d,k}\text{~and~} J(A_{d,k},\bs{X}_{-d,-k},\bs{w})\leq J(A_{d,k},\bs{X}_{-d,-k},\bs{w}')\nonumber\\
% &~\hspace{90mm}~\forall \bs{w}'\in\mathcal{W}\text{~and~}\bs{w}'\neq \bs{w}\Big\}.
% \end{align}
% Based on Lemma~\ref{lemma:propertyV} and the definition of $\phi_{\bs{w}}(A_{d,k})$,
Now, we have the following structural property for the optimal policy $\pi^*$.

\begin{theorem} \label{theorem:optimal}
If $\exists k\in\mathcal{K}$, such that $\bs{w}_k^*=(1,2)$, then $\pi^*(\bs{X})=\bs{w}^*$ for all $\bs{X}\in\mathcal{X}$ such that
% $ A_{d,k}\geq \phi_{\bs{w}^*}(\bs{X}_{-d,-k}).$
\begin{equation}
A_{d,k}\geq \phi_{\bs{w}^*}(\bs{X}_{-d,-k}).
\end{equation}
\end{theorem}
\begin{IEEEproof}
  See Appendix~\ref{app:optimal}.
  % See Appendix~B.
\end{IEEEproof}

From Theorem~\ref{theorem:optimal}, we observe that, for given $\bs{X}_{-d,-k}$, the scheduling action of $\bs{w}_k=(1,2)$ for device $k$ is threshold-based with respect to $A_{d,k}$. This indicates that, when  the AoI  $A_{d,k}$ of device $k$ is large,
it is more efficient for device $k$ to sample and transmit a new status update to the destination, as its previously sampled status update becomes rather obsolete and less valuable for the destination. Note that, different from most existing structural analysis solutions \cite{Koole} that typically require the monotonicity and multimodularity  of the value function, the structure in Theorem~\ref{theorem:optimal}  \emph{requires only the monotonicity of the value function} and the AoI dynamics in \eqref{eqn:aoi_device_k} and \eqref{eqn:aoi_bs_k}. Such a unique feature will be further exploited in Section~\ref{sec:suboptimal} to design a low-complexity suboptimal policy.
Theorem~\ref{theorem:optimal} implies that the optimal action for a certain system state is still optimal for some other system state.
% shows that, if it is optimal to schedule some devices to sample and start transmitting new status updates for some system state, then it is still optimal to choose this action for some other system states.
In particular, for all $\bs{X},\bs{X}'\in\mathcal{X}$, and $\bs{w}\in\mathcal{W}$ satisfying that $\bs{A}_r'=\bs{A}_r$, $\bs{D}'=\bs{D}$, and
\begin{equation}
\begin{cases}A_{d,k}'\geq A_{d,k},  & \text{if}~\bs{w}_k=(1,2), \\
            A_{d,k}'=A_{d,k},  &\text{otherwise},
  \end{cases},
\end{equation}
for all $k\in\mathcal{K}$, we have
\begin{align}\label{eqn:still_optimal}
\pi^*(\bs{X})=\bs{w}~\Rightarrow~\pi^*(\bs{X}')=\bs{w}.
\end{align}

The property in \eqref{eqn:still_optimal} can be leveraged to develop a low-complexity structure-aware relative value iteration algorithm and policy iteration algorithm, by extending their standard implementation. This can be done along the lines of the algorithm design in \cite{7915753}. These structure-aware optimal algorithms can use much less computational complexity compared to standard relative value iteration and policy iteration algorithms \cite{bertsekas4}.
% The computational saving particularly, when the number of devices is large.
However, they still suffer from the curse of dimensionality due to the exponential growth of the \textcolor{black}{state} space, i.e., $|\mathcal{X}|=\prod_{k\in\mathcal{K}}|\mathcal{X}_k|=\prod_{k\in\mathcal{K}}(\hat{A}_{d,k}+1)(\hat{A}_{r,k}+1)L_k$.
Thus, it is imperative to design low-complexity suboptimal solutions, by considering the structural properties of the optimal policy, as we do next.

\section{Low-Complexity Suboptimal Solution}\label{sec:suboptimal}
To overcome the curse of dimensionality, we propose a new, low-complexity suboptimal scheduling and sampling policy. We show that the structural property of the proposed policy is similar to that of the optimal policy. Then, we develop a new structure-aware algorithm to compute the proposed policy.
\subsection{Low-Complexity Suboptimal Policy}
The threshold structure of the optimal policy in Theorem~\ref{theorem:optimal} stems from the monotonicity of the value function.
Motivated by this, we apply a linear decomposition method for the value function, so that the monotonicity property can be maintained.
First, we introduce a semi-randomized base policy.

\begin{definition}\label{definition:semi-randomized-policy}
A \emph{semi-randomized scheduling and sampling base policy} is defined by $\hat{\pi}=(\hat{\pi}_{u},\hat{\pi}_{v})$, where $\hat{\pi}_{u}=(p_k^u)_{k\in\mathcal{K}}$ is a randomized scheduling policy, given by a distribution on the feasible scheduling action space $\mathcal{U}$ with $p_k^u\in[0,1]$ for each $k\in\mathcal{K}$ and $\sum_{k\in\mathcal{K}}p_k^u\leq M$,  and $\hat{\pi}_{v}$ is a deterministic sampling policy under a  given randomized  policy $\hat{\pi}_{u}$.
\end{definition}
% \footnote{Conditions $p_k^u\in[0,1]~\forall k$ and $\sum_{k\in\mathcal{K}}p_k^u\leq M$  are necessary and sufficient for the existence of a random scheduling policy under which no more than $K$ IoT devices are scheduled to transmit concurrently each time.}

% We restrict to unichain semi-randomized base policy.
Let $\hat{\theta}$ and $\hat{V}(\bs{X})$ be, respectively, the average \textcolor{black}{the AoI at the receiver} and the value function under a unichain semi-randomized base policy $\hat{\pi}$. Similar to Lemma~\ref{lemma:bellman}, there exists $(\theta, \hat{V}(\bs{X}))$
satisfying the following Bellman equation.
\begin{align}
  \hat{\theta} + \hat{V}(\bs{X}) =\sum_{k=1}^KA_{r,k}
  +  \min_{\bs{v}} \sum_{\bs{X}'\in\mathcal{X}}\mathbb{E}^{\hat{\pi}_{u}}\left[\Pr[\bs{X}'|\bs{X},\bs{w}]\right] V(\bs{X}'),~\forall \bs{X}\in\mathcal{X},\label{eqn:sub_bellman}
\end{align}
where $\Pr[\bs{X}'|\bs{X},\bs{w}]$ is given by \eqref{eqn:trans_prob}.
Next, we show that $\hat{V}(\bs{X})$ has the following additive separable structure.

\begin{lemma}\label{lemma:separable}
Given any unichain semi-randomized base policy $\hat{\pi}$, the value function $\hat{V}(\bs{X})$ in \eqref{eqn:sub_bellman} can be expressed as $\hat{V}(\bs{X})= \sum_{k\in\mathcal{K}} \hat{V}_k(\bs{X}_k)$, where for each $k$, $\hat{V}_k(\bs{X}_k)$ satisfies:
\begin{align}\label{eqn:per-bellman}
 \theta_k  + \hat{V}_k(\bs{X}_k) = A_{r,k}  + \min_{v_k} \sum_{\bs{X}_k'\in\mathcal{X}_k}\mathbb{E}^{\hat{\pi}_{u}}\left[\Pr[\bs{X}'_k|\bs{X}_k,\bs{w}_k]\right] \hat{V}_k(\bs{X}'_k),~\forall \bs{X}_k\in\mathcal{X}_k.
\end{align}
 Here, $\Pr[\bs{X}'_k|\bs{X}_k,\bs{w}_k]$ is given by \eqref{eqn:trans_prob_k}, $\theta_k$ and $\hat{V}_k(\bs{X}_k)$ are the per-device average \textcolor{black}{the AoI at the receiver} and the per-device value function under  policy $\hat{\pi}$, respectively.
\end{lemma}
\begin{IEEEproof}
  % See Appendix~\ref{app:seperable}.
  Along the line of the proof of \cite[Lemma 3]{harvest}, we prove the additive separable structure of the value function under a semi-randomized unichain base policy $\hat{\pi}$.
Due to the randomized scheduling action resulting from $\hat{\pi}_u$ and by making use of the relationship between the joint distribution and marginal distribution, we can obtain that, $ \sum_{\bs{X}'\in\mathcal{X}}\Pr[\bs{X}'|\bs{X},\bs{w}] = \sum_{\bs{X}_k'\in\mathcal{X}_k}\Pr[\bs{X}_k'|\bs{X},\bs{w}] = \sum_{\bs{X}_k'\in\mathcal{X}_k}\Pr[\bs{X}_k'|\bs{X}_k,\bs{w}_k]$
% \begin{align}
% \sum_{\bs{X}'\in\mathcal{X}}\Pr[\bs{X}'|\bs{X},\bs{w}] = \sum_{\bs{X}_k'\in\mathcal{X}_k}\Pr[\bs{X}_k'|\bs{X},\bs{w}] = \sum_{\bs{X}_k'\in\mathcal{X}_k}\Pr[\bs{X}_k'|\bs{X}_k,\bs{w}_k]
% \end{align}
holds for each state $\bs{X}$ and the semi-randomized control action $\bs{w}=\hat{\pi}_u(\bs{X})$.
Then, by substituting $\hat{V}(\bs{X})=\sum_{k\in\mathcal{K}}\hat{V}_k(\bs{X}_k)$ into \eqref{eqn:sub_bellman}, it can be easily checked that the equality in \eqref{eqn:per-bellman} holds. We complete the proof.
  % See Appendix~C.
\end{IEEEproof}
% Based on Lemma~\ref{lemma:separable},

Now, we approximate the value function in \eqref{eqn:bellman} with $\hat{V}(\bs{X})$:
% \begin{equation}
% V(\bs{X})\approx\hat{V}(\bs{X}) = \sum_{k\in\mathcal{K}} \hat{V}_k(\bs{X}_k),
% \end{equation}
$V(\bs{X})\approx\hat{V}(\bs{X}) = \sum_{k\in\mathcal{K}} \hat{V}_k(\bs{X}_k),$
where $\hat{V}_k(\bs{X}_k)$ is given by \eqref{eqn:per-bellman}.
Then, according to \eqref{eqn:optimal_pi}, we develop a deterministic scheduling and sampling suboptimal policy $\hat{\pi}^*$ as follows.
\begin{align}
  \hat{\pi}^*(\bs{X})=\arg\min_{\bs{w}\in\mathcal{W}}\sum_{\bs{X}'\in\mathcal{X}}\Pr[\bs{X}'|\bs{X},\bs{w}] \sum_{k\in\mathcal{K}}\hat{V}_k(\bs{X}'_k),~\forall \bs{X}\in\mathcal{X}.\label{eqn:suboptimal_pi}
\end{align}

\textcolor{black}{ The proposed deterministic policy $\hat{\pi}^*$ in \eqref{eqn:suboptimal_pi} resembles the one iteration step in the standard policy iteration algorithm. By making use of the proof in establishing the convergence of policy iteration (i.e., the monotonicity of the iterations of policy iteration), e.g., \cite[Proposition 5.4.2]{bertsekas4} and \cite[Theorem 8.6.6]{puterman}, and following arguments similar to those used in proving \cite[Theorem 1]{harvest}, we can then state the proposed deterministic policy $\hat{\pi}^*$ will always outperform the corresponding semi-randomized base policy $\hat{\pi}$.}

The computational complexity needed for obtaining the proposed policy $\hat{\pi}^*$ is much lower than the one needed for the optimal policy $\pi^*$ in  \eqref{eqn:optimal_pi}.
In particular, to obtain the proposed policy $\hat{\pi}^*$, we need to compute $\{\hat{V}_k(\bs{X}_k)\}$ for each device $k$, which is a total of $O(\sum_{k\in\mathcal{K}}(\hat{A}_{d,k}+1)(\hat{A}_{r,k}+1)L_k)$ values.
In contrast, obtaining the optimal policy $\pi^*$ by computing $\{V(\bs{X})\}$ requires a total of $O(\prod_{k\in\mathcal{K}}(\hat{A}_{d,k}+1)(\hat{A}_{r,k}+1)L_k)$ values.
Thus, the complexity needed to compute $\hat{\pi}^*$ decreases from exponential with $K$ to linear with $K$.

\subsection{Structural Analysis and Algorithm Design}

Now, we investigate the structural properties of the proposed suboptimal policy $\hat{\pi}^*$. First, we show the following property of the per-device value function $\hat{V}_k(\bs{X}_k)$, for a given semi-randomized base policy $\hat{\pi}$.

\begin{lemma}\label{lemma:propertyVk}
Given a semi-randomized base policy $\hat{\pi}$, for all $k\in\mathcal{K}$, we have $\hat{V}_k(\bs{X}_k^2)\geq \hat{V}_k(\bs{X}_k^1)$ for any $\bs{X}_k^1,\bs{X}_k^2 \in\mathcal{X}_k$ such that $A_{d,k}^2 \geq A_{d,k}^1$, $A_{r,k}^2 \geq A_{r,k}^1$, and $D_k^2=D_k^1$.
\end{lemma}
\begin{IEEEproof}
  See Appendix~\ref{app:per-value-function}.
  % See Appendix~C.
\end{IEEEproof}

Similar to the  analysis for the optimal policy, we define:
\begin{align*}%\label{eqn:sub_threshold}
\hat{\phi}_{\bs{w}}(\bs{X}_{-d,-k})\triangleq\begin{cases}\min\hat{\Phi}_{\bs{w}}(\bs{X}_{-d,-k}),  & \text{if}~\hat{\Phi}_{\bs{w}}(\bs{X}_{-d,-k})\neq\emptyset, \\
            +\infty,  &\text{otherwise},
  \end{cases},
\end{align*}
where $\hat{\Phi}_{\bs{w}}(\bs{X}_{-d,-k}) \triangleq \{ A_{d,k}| A_{d,k}\in\mathcal{A}_{d,k}\text{~and~} \hat{J}(A_{d,k},\bs{X}_{-d,-k},\bs{w})  \leq  \hat{J}(A_{d,k},\bs{X}_{-d,-k},\bs{w}')~\forall \bs{w}'\in\mathcal{W}\text{~and~}\bs{w}'\neq \bs{w}\}$
 and $\hat{J}(\bs{X},\bs{w})\triangleq\sum_{k=1}^KA_{r,k} + \sum_{\bs{X}'\in\mathcal{X}}\Pr[\bs{X}'|\bs{X},\bs{w}] \sum_{k\in\mathcal{K}}\hat{V}_k(\bs{X}'_k)$.
Then, we can show the structural property of the proposed policy $\hat{\pi}^*$.
\begin{theorem} \label{theorem:suboptimal}
If $\exists k\in\mathcal{K}$, such that $\hat{\bs{w}}_k^*=(1,2)$, then $\hat{\pi}^*(\bs{X})=\hat{\bs{w}}^*$ for all $\bs{X}\in\mathcal{X}$ such that %$A_{d,k}\geq \hat{\phi}_{\hat{\bs{w}}^*}(\bs{X}_{-d,-k})$.
\begin{equation}
A_{d,k}\geq \hat{\phi}_{\hat{\bs{w}}^*}(\bs{X}_{-d,-k}).
\end{equation}
\end{theorem}
\begin{IEEEproof}
  See Appendix~\ref{app:suboptimal}.
  % See Appendix~C.
\end{IEEEproof}

By comparing Theorem~\ref{theorem:suboptimal} with Theorem~\ref{theorem:optimal}, We can see that the proposed policy $\hat{\pi}^*$ possesses a threshold-based structure similar to the optimal policy $\pi^*$. This is mainly due to the linear decomposition method and  the special properties of the AoI dynamics in \eqref{eqn:aoi_device_k} and \eqref{eqn:aoi_bs_k}.
 % Such  similarity provides a structural base for the good performance of the proposed policy, as will be shown in the simulations.

Theorem~\ref{theorem:suboptimal} exhibits a similar property to \eqref{eqn:still_optimal}. Thus, in Algorithm~\ref{alg:suboptimal}, we propose a structure-aware algorithm to compute the suboptimal policy by making use of its structure. Note that, whenever the ``if'' condition in Algorithm~\ref{alg:suboptimal} is satisfied for certain system states, then we can immediately obtain the corresponding control action, without performing the minimization in \eqref{eqn:suboptimal_pi}. This yields  considerable computational saving, particularly, for a large number of devices, i.e., a large $\mathcal{W}$.

\begin{algorithm}[!t]
\caption{Proposed Low-Complexity Structure-Aware Algorithm}
\label{alg:suboptimal}
\begin{algorithmic}[1]
\State Given a semi-randomized base policy $\hat{\pi}$, for each $k\in\mathcal{K}$, compute the per-device value function $\{\hat{V}_k(\bs{X}'_k)\}$ for all $\bs{X}_k\in\mathcal{X}_k$ by solving the Bellman equation in \eqref{eqn:per-bellman}.

\State Obtain the proposed deterministic policy $\hat{\pi}^*$ for each $\bs{X}$:

\Statex\textbf{if} $\exists k\in\mathcal{K}$ and $\bs{X}'\in\mathcal{X}$ such that $\hat{\pi}^*(\bs{X}') = \bs{w}$ with $\bs{w}_k=(1,2)$, $\bs{A}_r'=\bs{A}_r$, $\bs{D}'=\bs{D}$, $A_{d,j}'\leq A_{d,j}$ for $j=k$ and $A_{d,j}= A_{d,j}'$ for $j\neq k$, \textbf{then}

\hspace{20mm} $\hat{\pi}^*(\bs{X}) = \bs{w}$.

\Statex\textbf{else}

\hspace{20mm} Compute $\hat{\pi}^*(\bs{X})$ by \eqref{eqn:suboptimal_pi}.

\Statex\textbf{end if}

\end{algorithmic}
\end{algorithm}

\section{IoT Monitoring System with Random Status Updates Arrivals}\label{sec:extension}
\begin{figure}[!t]
\begin{centering}
\includegraphics[scale=.44]{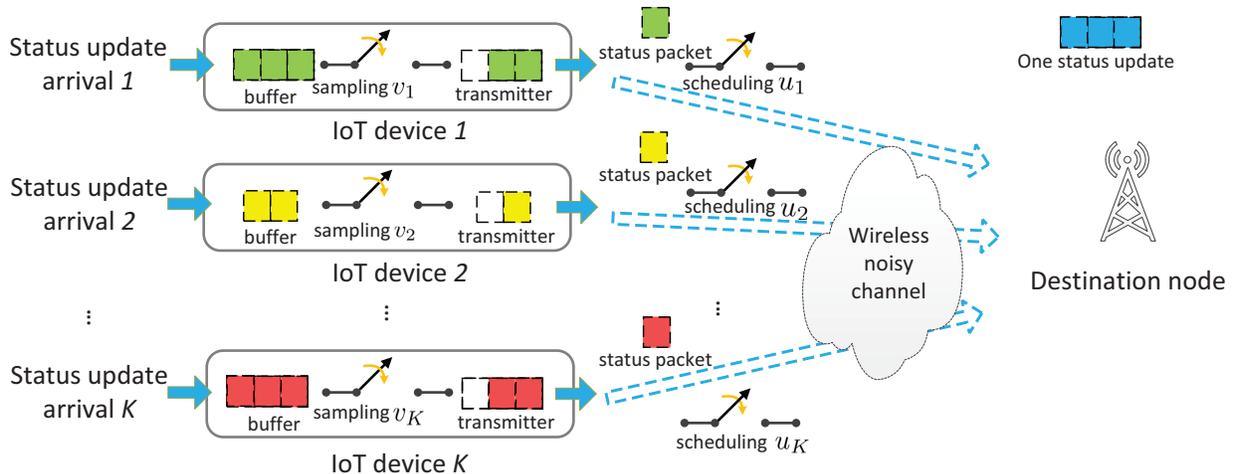}
 \caption{Illustration of a real-time IoT monitoring system with random status update arrivals.}\label{fig:system-extension}
\end{centering}
 % \vspace{-0.5cm}
\end{figure}

Thus far, we have studied the optimal device scheduling and status sampling control for a real-time IoT monitoring system, where the status information updates can be generated at will by each IoT device.
Now, we extend the monitoring system in Section~\ref{sec:systemmodel} to an IoT system in which the status information updates arrive at each IoT device randomly and are queued at each IoT device before being transmitted to the destination, as illustrated in Fig.~\ref{fig:system-extension}. 
\textcolor{black}{Note that such a scenario is not considered in our previous work \cite{8778671}.}

\subsection{Scheduling with Random Status Updates}
We still consider a discrete-time system with slots indexed by $t=1,2,\cdots$ and the status update (if any) arrives at each IoT device at the beginning of each time slot.
Similar to \cite{HsuISIT,jiang2018can,Talak}, we assume that the status update arrivals for different IoT devices are  mutually independent, and for each IoT device $k\in\mathcal{K}$, the status update arrivals are independent and identically distributed (i.i.d.) over time slots, following a Bernoulli distribution with mean rate $\rho_k\in[0,1]$.
We assume that each IoT device is equipped with a buffer to store the newly arriving status update, as in \cite{jiang2018can} and \cite{Talak}.
\textcolor{black}{We consider that the current in-transmission status update is not stored in the buffer, and thus, will not be replaced by a newly arriving status update.}
We consider that the newly arrived status update, i.e., the most recent update, will replace the older one (if any) in the buffer of each IoT device, as the destination will not benefit from receiving an outdated status update.
The models of the non-uniform status packet sizes and the noisy channels are similar to those in Section~\ref{sec:systemmodel}.

In each slot, the network also needs to determine which IoT devices to  schedule so as to update their status. The scheduling action $u_k(t)\in\{0,1\}$ for each IoT device $k$ remains the same as in the deterministic case.
However, due to the random status update arrivals, the sampling control action of each IoT device will be different.
Specifically, for each scheduled device, if there is no status packet stored in its buffer, then, the network will schedule this device to continue with its current in-transmission update,
otherwise, the network must decide whether to continue the current in-transmission update  or start to transmit the status update in the buffer.
With some notation abuse, let $v_k\in\{1,2\}$ be the sampling action for each IoT device $k$, where $v_k=1$ indicates that device $k$ will continue transmitting its current in-transmission update, and $v_k=2$ indicates that device $k$ will start transmitting the status update in its buffer and drop the current in-transmission update.
Accordingly, the system control action at slot $t$ is denoted as $\bs{w}(t)= (\bs{u}(t),\bs{v}(t))\in\mathcal{W}\triangleq\mathcal{U}\times\mathcal{V}$, where $\bs{u}(t)=(u_k(t))_{k\in\mathcal{K}}\in\mathcal{U}$ is the system scheduling action and $\bs{v}(t)=(v_k(t))_{k\in\mathcal{K}}\in\mathcal{V}$ is the system sampling action.

% Next, we introduce the AoI model for the considered system.
Due to the buffer at each device $k$, % except for the AoI $A_{d,k}$ at device $k$ , the AoI $A_{r,k}$ of the destination for device $k$, and the number  $D_k$ of remaining packets of the current in-transmission update,
except for $A_{d,k}$, $A_{r,k}$, and $D_k$, we need to further introduce the age of the status update in the buffer, which is referred to as the AoI at the buffer at device $k$.
We denote by $A_{b,k}(t)\in\mathcal{A}_{b,k}$ the AoI at the buffer at device $k$ at the beginning of slot $t$, where $\mathcal{A}_{b,k}\triangleq \{0,1,\cdots,\hat{A}_{b,k}\}$ is the state space for the AoI at the buffer at device $k$ and $\hat{A}_{b,k}$ is the corresponding upper limit. We also assume that $\hat{A}_{b,k}$ is finite, but can be arbitrarily large.
With some abuse of notation, let $\bs{X}_k(t) \triangleq (A_{b,k}(t), A_{d,k}(t),A_{r,k}(t),D_k(t))\in\mathcal{X}_k\triangleq \mathcal{A}_{b,k}\times\mathcal{A}_{d,k}\times \mathcal{A}_{r,k}\times \mathcal{D}_{k}$ be the system state vector of device $k$ at slot $t$ and let $\bs{X}(t)\triangleq (\bs{X}_k(t))_{k\in\mathcal{K}}\in\mathcal{X}\triangleq \prod_{k\in\mathcal{K}}\mathcal{X}_k$ be the system state matrix at slot $t$.

Next, we study how $\bs{X}_k(t)$ evolves with the system control action $\bs{w}_k(t)$. Note that, \textcolor{black}{the AoI at the receiver} depends on the AoI at each device, which depends on the AoI at the buffer at each device. It can be seen that, the dynamics of $A_{r,k}(t)$ and $D_k(t)$ are the same to those in Section~\ref{sec:aoi model}, given by \eqref{eqn:aoi_bs_k} and \eqref{eqn:d_device_k}, respectively.
For the AoI at the buffer at device $k$, if there is a status update arriving at device $k$ at slot $t$, then the AoI will decrease to one, otherwise, the AoI will increase by one. As a result, the AoI dynamics of the buffer at device $k$ are given by:
\begin{align}\label{eqn:aoi_buffer_k_extension}
A_{b,k}(t+1)
&=\begin{cases} 1,  &\text{if status update arrives at $t$},\\
                \min\{A_{b,k}(t)+1,\hat{A}_{b,k}\}, &\text{otherwise.}
      \end{cases}
\end{align}

For the AoI at device $k$, when device $k$ is scheduled to continue sending its current in-transmission update at slot $t$ (i.e., $\bs{w}_k(t)=(1,1)$), if there remains only one packet and the transmission is successful, then the AoI will decrease to the AoI at the buffer at device $k$ at slot $(t+1)$. When device $k$ is scheduled to transmit the status update in its buffer at $t$ (i.e., $\bs{w}_k(t)=(1,2)$), then the AoI will decrease to the AoI at the buffer at device $k$ at slot $t$ plus one, irrespective of whether the transmission is successful or not. In all other cases, the AoI will increase by one. Thus, the AoI dynamics of device $k$ are given by:
\begin{align}\label{eqn:aoi_device_k_extension}
A_{d,k}(t+1)
&=\begin{cases} &\min\{A_{b,k}(t+1),\hat{A}_{d,k}\},  ~\text{if}~\bs{w}_k(t)=(1,1),D_k(t)=1,\text{and}\\
&\hspace{43mm}\text{transmission succeeds at}~t,\\
&\min\{A_{b,k}(t)+1,\hat{A}_{d,k}\}, ~ \text{if}~ \bs{w}_k(t)=(1,2),\\
                &\min\{A_{d,k}(t)+1,\hat{A}_{d,k}\}, ~ \text{otherwise.}
      \end{cases}
\end{align}

\subsection{Problem Formulation}
Similar to Section~\ref{sec:problem formulation}, given an observed system state $\bs{X}$, the system scheduling and sampling action $\bs{w}$ is derived according to a feasible stationary scheduling policy $\pi=(\pi_{u},\pi_{v})$, which is defined in the same manner as Definition~\ref{definition:stationary_policy_IoT}. Following the dynamics in \eqref{eqn:d_device_k}, \eqref{eqn:aoi_bs_k}, \eqref{eqn:aoi_buffer_k_extension}, and \eqref{eqn:aoi_device_k_extension}, the induced random process $\bs{X}(t)$ for a given policy $\pi$ is a controlled Markov chain with the following transition probability:

\begin{equation}
\Pr[\bs{X}'|\bs{X},\bs{w}] = \prod_{k=1}^K \Pr[\bs{X}'_k|\bs{X}_k,\bs{w}_k],\label{eqn:trans_prob_extension}
\end{equation}
where
\begin{align}\label{eqn:trans_prob_k_extension}
&\Pr[\bs{X}'_k|\bs{X}_k,\bs{w}_k]\nonumber\\
&=\Pr[\bs{X}_k(t+1)=\bs{X}'_k|\bs{X}_k=\bs{X}_k(t),\bs{w}_k=\bs{w}_k(t)]\nonumber\\
&=\begin{cases}
\rho_k\lambda_k,  &~\text{if}~\bs{X}'_k = \bs{X}_{k,s}^1~\text{and~} u_k=1,\\
\rho_k(1-\lambda_k), &~\text{if}~\bs{X}'_k = \bs{X}_{k,f}^1~\text{and~} u_k=1,\\
(1-\rho_k)\lambda_k,  &~\text{if}~\bs{X}'_k = \bs{X}_{k,s}^2~\text{and~} u_k=1,\\
(1-\rho_k)(1-\lambda_k), &~\text{if}~\bs{X}'_k = \bs{X}_{k,f}^2~\text{and~} u_k=1,\\
 \rho_k,&~\text{if}~\bs{X}'_k = \bs{X}_{k,un}^1~\text{and~} u_k=0,\\
 (1-\rho_k),&~\text{if}~\bs{X}'_k = \bs{X}_{k,un}^2~\text{and~} u_k=0,\\
                0, &~\text{otherwise.}
      \end{cases}
\end{align}
Here, $\bs{X}_{k,s}^1$, $\bs{X}_{k,f}^1, \bs{X}_{k,un}^1$ and $\bs{X}_{k,s}^2$, $\bs{X}_{k,f}^2, \bs{X}_{k,un}^2$ indicate whether a new status update arrives at the device or not, and $\bs{X}_{k,s}^1$, $\bs{X}_{k,s}^2$ and $\bs{X}_{k,f}^1$, $\bs{X}_{k,f}^2$ indicate whether a transmission succeeds or fails.
According to \eqref{eqn:d_device_k}, \eqref{eqn:aoi_bs_k}, \eqref{eqn:aoi_buffer_k_extension}, and \eqref{eqn:aoi_device_k_extension}, we know that, if $v_k=1$, i.e., device $k$ is scheduled to continue sending its current in-transmission update, then
\begin{align}
&\bs{X}_{k,s}^1
=\begin{cases}
(1,1,\min\{A_{d,k}+1,\hat{A}_{r,k}\}, L_k),  &\text{if}~D_k=1,\\
(1,\min\{A_{d,k}+1,\hat{A}_{d,k}\},\min\{A_{r,k}+1,\hat{A}_{r,k}\}, D_k-1),
&\text{otherwise.}
      \end{cases}\\
&\bs{X}_{k,s}^2
=\begin{cases}
(\min\{A_{b,k}+1,\hat{A}_{b,k}\},\min\{A_{b,k}+1,\hat{A}_{d,k}\},\min\{A_{d,k}+1,\hat{A}_{r,k}\}, L_k),  &\text{if}~D_k=1,\\
(\min\{A_{b,k}+1,\hat{A}_{b,k}\},\min\{A_{d,k}+1,\hat{A}_{d,k}\},\min\{A_{r,k}+1,\hat{A}_{r,k}\}, D_k-1),
&\text{otherwise.}
      \end{cases}\\
 &\bs{X}_{k,f}^1= (1,\min\{A_{d,k}+1,\hat{A}_{d,k}\},\min\{A_{r,k}+1,\hat{A}_{r,k}\}, D_k),\\
 &\bs{X}_{k,f}^2= (\min\{A_{b,k}+1,\hat{A}_{b,k}\},\min\{A_{d,k}+1,\hat{A}_{d,k}\},\min\{A_{r,k}+1,\hat{A}_{r,k}\}, D_k);
\end{align}
% \begin{equation}
% \bs{X}_{k,f}= (\min\{A_{d,k}+1,\hat{A}_{d,k}\},\min\{A_{r,k}+1,\hat{A}_{r,k}\}, D_k).
% \end{equation}
if $v_k=2$, i.e., device $k$ is scheduled to start sending the status update in its buffer, then
\begin{align}
&\bs{X}_{k,s}^1=(1,1,\min\{A_{r,k}+1,\hat{A}_{r,k}\}, L_k-1),\\
&\bs{X}_{k,s}^2=(\min\{A_{b,k}+1,\hat{A}_{b,k}\},\min\{A_{b,k}+1,\hat{A}_{d,k}\},\min\{A_{r,k}+1,\hat{A}_{r,k}\}, L_k-1),\\
&\bs{X}_{k,f}^1= (1,1,\min\{A_{r,k}+1,\hat{A}_{r,k}\}, L_k),\\
&\bs{X}_{k,f}^2= (\min\{A_{b,k}+1,\hat{A}_{b,k}\},\min\{A_{b,k}+1,\hat{A}_{d,k}\},\min\{A_{r,k}+1,\hat{A}_{r,k}\}, L_k);
\end{align}
 and %$\bs{X}_{k,un}= (\min\{A_{d,k}+1,\hat{A}_{d,k}\},\min\{A_{r,k}+1,\hat{A}_{r,k}\}, D_k)$.
\begin{align}
&\bs{X}_{k,un}^1= (1,\min\{A_{d,k}+1,\hat{A}_{d,k}\},\min\{A_{r,k}+1,\hat{A}_{r,k}\}, D_k).\\
&\bs{X}_{k,un}^2= (\min\{A_{b,k}+1,\hat{A}_{b,k}\},\min\{A_{d,k}+1,\hat{A}_{d,k}\},\min\{A_{r,k}+1,\hat{A}_{r,k}\}, D_k).
\end{align}

Then, as before, we aim to find the optimal feasible stationary unichain scheduling and sampling policy that minimizes the average \textcolor{black}{the AoI at the receiver}, given by:
\begin{align}
\bar{A}_r^*(\bs{X}_1)\triangleq\min_{\pi}\bar{A}_r^{\pi}(\bs{X}_1)\triangleq\limsup_{T\to\infty}\frac{1}{T}\sum_{t=1}^T \sum_{k=1}^K \mathbb{E} \left[A_{r,k}(t)|\bs{X}_1\right].\label{eqn:mdp-extension}
\end{align}
Similar to Lemma~\ref{lemma:bellman}, the optimal policy $\pi^*$ can be obtained by solving the corresponding Bellman equation, given by:
\begin{align}
  \theta + V(\bs{X}) = \sum_{k=1}^KA_{r,k} + \min_{\bs{w}\in\mathcal{W}} \sum_{\bs{X}'\in\mathcal{X}}\Pr[\bs{X}'|\bs{X},\bs{w}] V(\bs{X}'),~\forall \bs{X}\in\mathcal{X},\label{eqn:bellman-extension}
\end{align}
where $\Pr[\bs{X}'|\bs{X},\bs{w}]$ is given by \eqref{eqn:trans_prob_extension}.

\subsection{Structural Properties of the Optimal Policy}
Following the line of the analysis in Section~\ref{sec:optimal policy}, we characterize the structural properties of the optimal scheduling and sampling policy $\pi^*$ for the MDP in \eqref{eqn:mdp-extension}.
First, we show that the monotonicity property of the value function $V(\bs{X})$. Define $\bs{A}_b\triangleq (A_{b,k})_{k\in\mathcal{K}}$.
\begin{lemma}\label{lemma:propertyV_extension}
For any $\bs{X}^1,\bs{X}^2 \in\mathcal{X}$ such that $\bs{A}_b^2\succeq\bs{A}_b^1$, $\bs{A}_d^2\succeq\bs{A}_d^1$, $\bs{A}_r^2\succeq\bs{A}_r^1$, and $\bs{D}^2=\bs{D}^1$, we have $V(\bs{X}^2)\geq V(\bs{X}^1)$.
\end{lemma}
% \begin{IEEEproof}
%   See Appendix~\ref{app:propertyV_extension}.
%   % See Appendix~A.
% \end{IEEEproof}

The proof is similar to the proof for Lemma~\ref{lemma:propertyV} in Appendix~\ref{app:propertyV}, and thus, is omitted here.
From Lemma~\ref{lemma:propertyV_extension}, we can see that, for the considered system with random status update arrivals, the value function $V(\bs{X})$ for the MDP in \eqref{eqn:mdp-extension} is also non-decreasing with the AoI $A_{b,k}$ of the buffer of each IoT device.  Then, we introduce the state-action cost function $J(\bs{X},\bs{w})$ and the function $\phi_{\bs{w}}(\bs{X}_{-d,-k})$ in the same manner as \eqref{eqn:J_function} and  \eqref{eqn:threshold}, respectively. Now, following the proof for Theorem~\ref{theorem:optimal} in Appendix~\ref{app:optimal}, we can show the following structural property for $\pi^*$.
\begin{theorem} \label{theorem:optimal_extension}
If $\exists k\in\mathcal{K}$, such that $\bs{w}_k^*=(1,2)$, then $\pi^*(\bs{X})=\bs{w}^*$ for all $\bs{X}\in\mathcal{X}$ such that
% $ A_{d,k}\geq \phi_{\bs{w}^*}(\bs{X}_{-d,-k}).$
\begin{equation}
A_{d,k}\geq \phi_{\bs{w}^*}(\bs{X}_{-d,-k}).
\end{equation}
\end{theorem}
% \begin{IEEEproof}
%   See Appendix~\ref{app:optimal_extension}.
%   % See Appendix~B.
% \end{IEEEproof}

From Theorem~\ref{theorem:optimal_extension}, we can see that, the structure of the optimal policy is very similar to the one in Theorem~\ref{theorem:optimal}. 
\textcolor{black}{Note that, for Theorem~\ref{theorem:optimal_extension}, the considered MDP is substantially different from the MDP for the case in Section~\ref{sec:systemmodel}, due to different dynamics of system states and transition probabilities. Moreover, the system state $\bs{X}$, which consists of the AoI at the buffer at each device, is also different from the system state used in Theorem~\ref{theorem:optimal}.}
% However, note that, we have a different definition of the system state $\bs{X}$ here, which consists of the AoI at the buffer at each device.
Theorem~\ref{theorem:optimal_extension} indicates that the scheduling action of $\bs{w}_k=(1,2)$ is threshold-based with $A_{d,k}$, for given $\bs{X}_{-d,-k}$. Similar arguments on the insights of such structure for Theorem~\ref{theorem:optimal} in Section~\ref{sec:optimal policy} can be drawn here.
Theorem~\ref{theorem:optimal_extension} indicates that, for all $\bs{X},\bs{X}'\in\mathcal{X}$, and $\bs{w}\in\mathcal{W}$ satisfying that $\bs{A}_b'=\bs{A}_b$, $\bs{A}_r'=\bs{A}_r$, $\bs{D}'=\bs{D}$, and
\begin{equation}
\begin{cases}A_{d,k}'\geq A_{d,k},  & \text{if}~\bs{w}_k=(1,2), \\
            A_{d,k}'=A_{d,k},  &\text{otherwise},
  \end{cases},
\end{equation}
for all $k\in\mathcal{K}$, we have
\begin{align}\label{eqn:still_optimal_extension}
\pi^*(\bs{X})=\bs{w}~\Rightarrow~\pi^*(\bs{X}')=\bs{w}.
\end{align}
Along the lines of the algorithm design in Section V, we can also exploit the structural property in \eqref{eqn:still_optimal_extension}  to develop a structure-aware low-complexity suboptimal solution.

\section{Simulation Results and Analysis}
In this section, we present numerical results to illustrate the structure of the optimal policies in Sections~\ref{sec:optimal policy} and \ref{sec:extension},  and the performance of the proposed suboptimal policy in Section~\ref{sec:suboptimal}.
Here, for the semi-randomized base policy $\hat{\pi}$, we consider that the probability $p_{k}^u$ of scheduling device $k$  is proportional to its channel reliability $\lambda_k$, i.e., $p_{k}^u = \lambda_k\slash \sum_{j}\lambda_j$ for $k\in\mathcal{K}$.
\textcolor{black}{We consider a greedy baseline policy, in which, the scheduling policy is determined by choosing the top $M$ users with the highest AoI at the receiver $A_{r,k}$ and the sampling policy is determined by solving a per-device Bellman equation for each device in a similar way to \eqref{eqn:per-bellman}.}
% We set the upper limits of the AoI at device $k$ and \textcolor{black}{the AoI at the receiver} for device $k$ such that $\hat{A}_{d,k}=\hat{A}_{r,k}=10$ for all $k$, unless stated otherwise.
% We set $\hat{A}_{d,k}$$=$$\hat{A}_{r,k}$$=$$10$ for all $k$.
 % We set the upper limits of the AoI at device $k$ and \textcolor{black}{the AoI at the receiver} for device $k$, $\hat{A}_{d,k}$ and $\hat{A}_{r,k}$, be $10$ for all $k$.

\begin{figure}[!t]
\begin{minipage}[c]{0.475\linewidth}
\centering
       \includegraphics[scale=0.47]{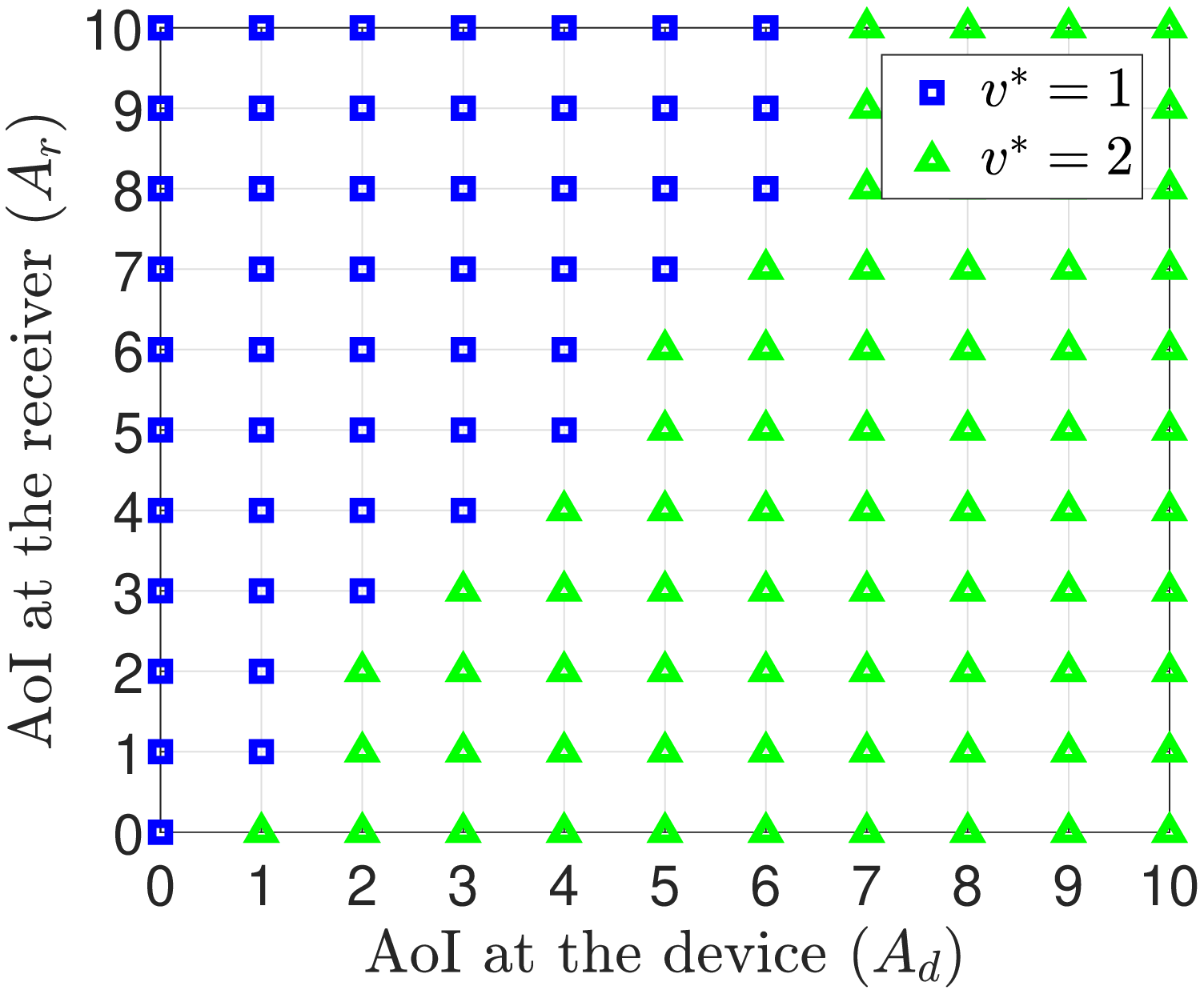}
\subcaption{}\label{fig:single_D1}
\end{minipage}
\begin{minipage}[c]{0.475\linewidth}
\centering
        \includegraphics[scale=0.47]{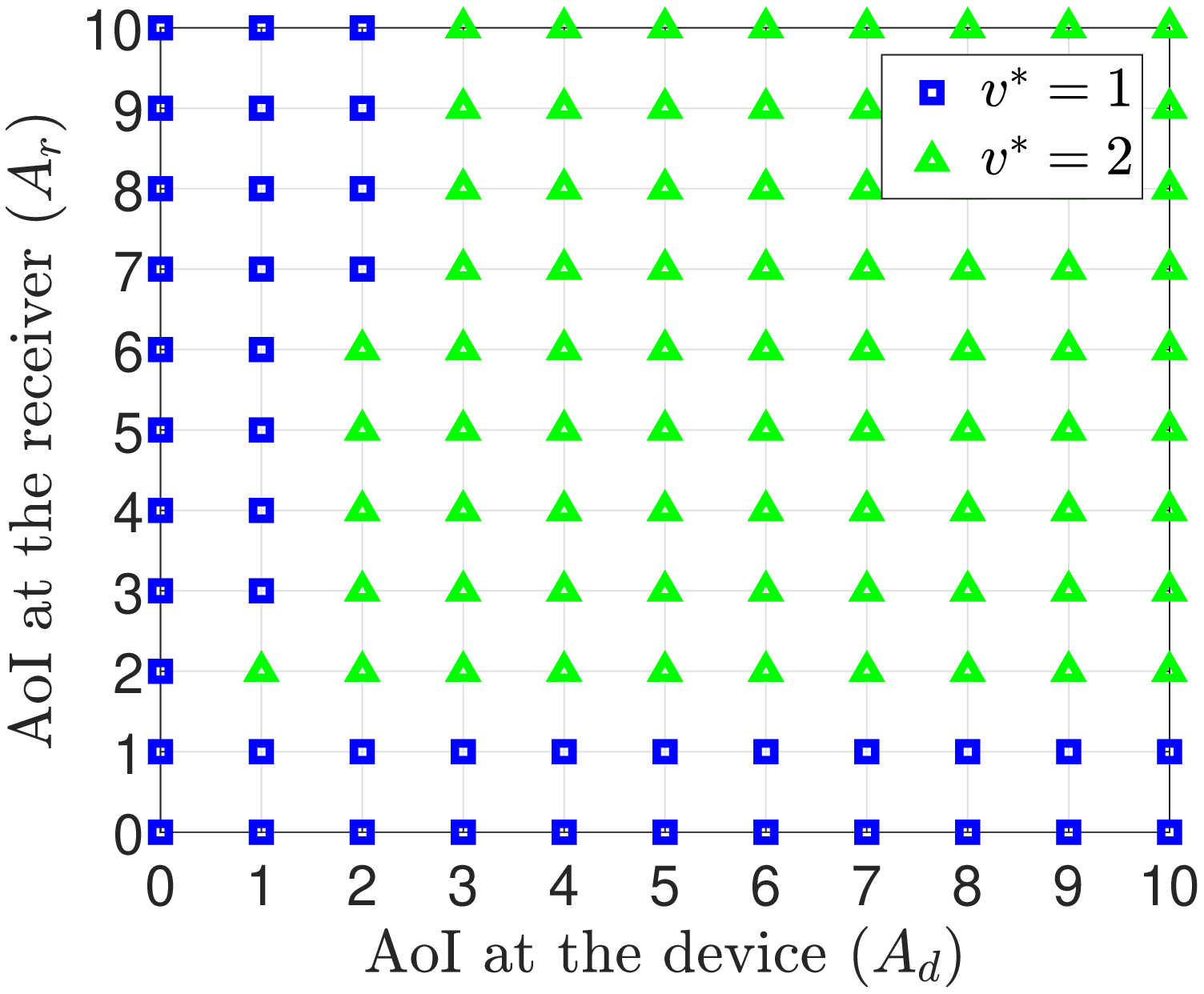}
\subcaption{}\label{fig:single_D3}
\end{minipage}
% \vspace{-0.3cm}
\caption{Structure of optimal policy $\pi^*$ in the single IoT device case. $\hat{A}_{l}=\hat{A}_{r}=10$, $L=4$, and $\lambda = 0.8$. (a) $D=1$. (b) $D=3$.}\label{fig:single}
 % \vspace{-0.6cm}
\end{figure}

\begin{figure}[!t]
\begin{minipage}[h]{\linewidth}
\centering
       \includegraphics[scale=0.47]{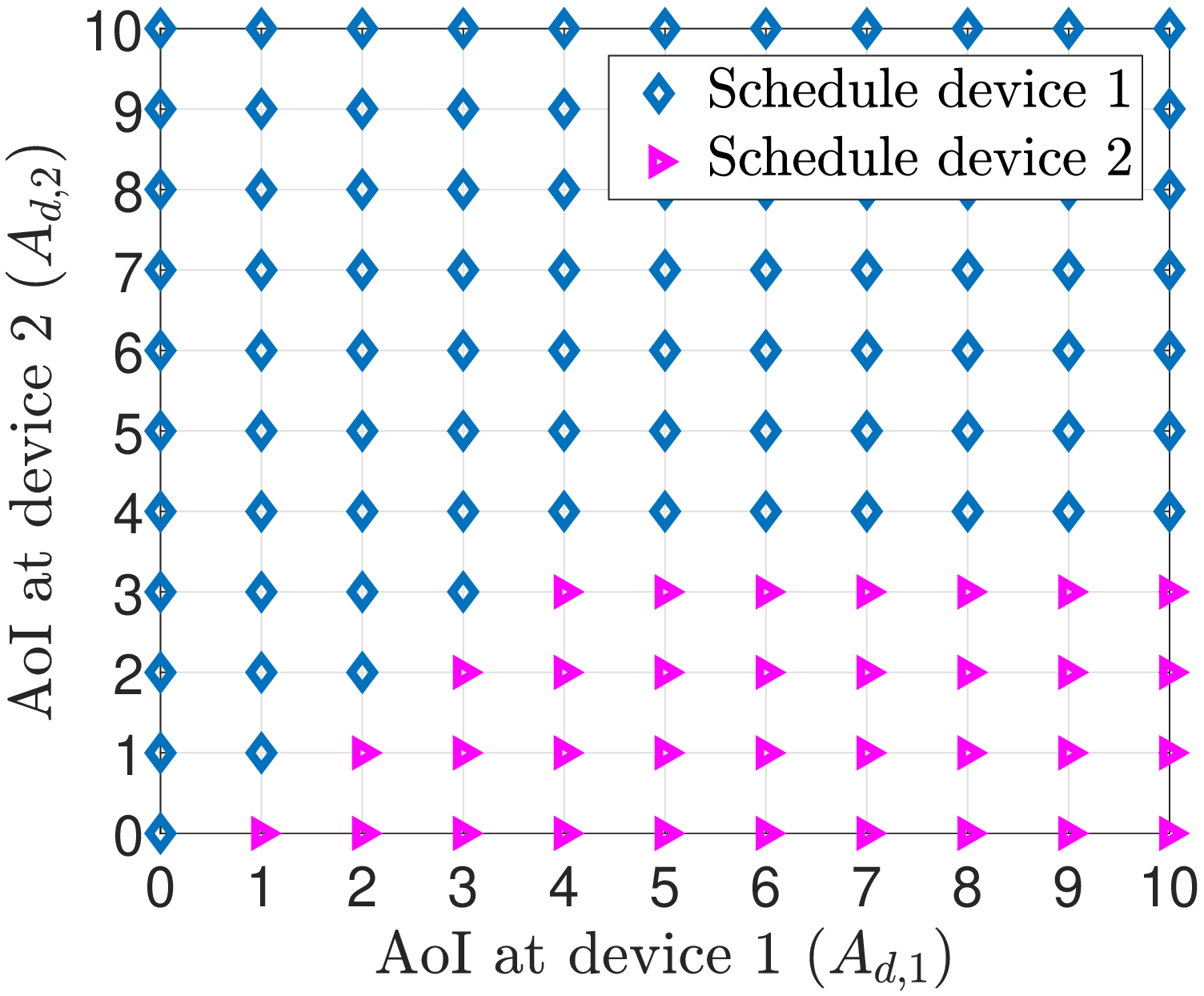}
\subcaption{}\label{fig:two_equal}
\end{minipage}
\begin{minipage}[h]{.475\linewidth}
\centering
        \includegraphics[scale=0.47]{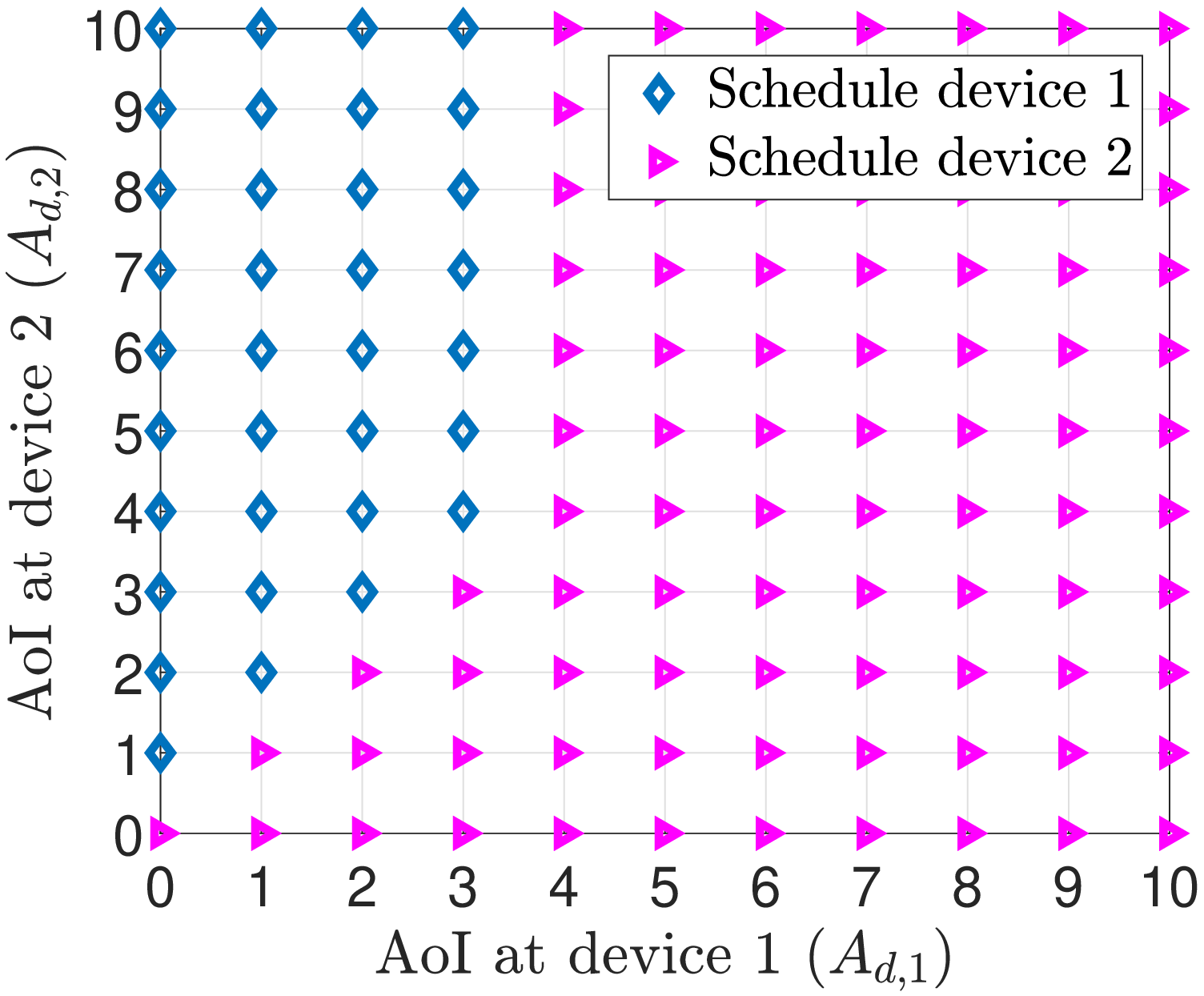}
\subcaption{}\label{fig:two_higher}
\end{minipage}
\begin{minipage}[h]{.475\linewidth}
\centering
        \includegraphics[scale=0.47]{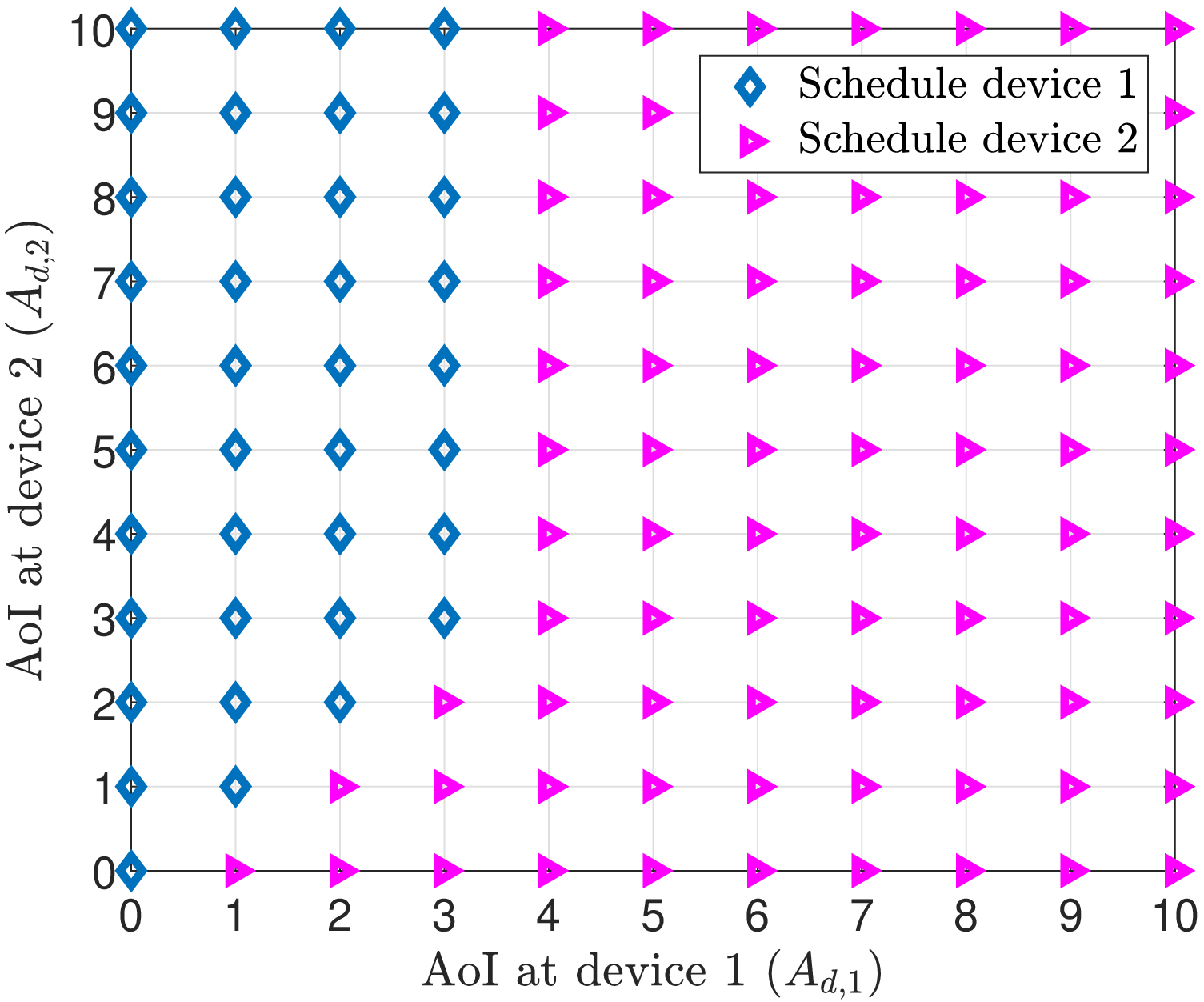}
\subcaption{}\label{fig:two_D_heter}
\end{minipage}
% \vspace{-0.2cm}
\caption{Structure of optimal policy $\pi^*$ for a case of two IoT device. $\hat{A}_{d,k}=\hat{A}_{r,k}=10$, for $k=1,2$. $A_{r,1}=A_{r,2}=5$, $D_1=D_2=1$, and $M=1$. (a) $\lambda_1=\lambda_2=0.7$ and $L_1=L_2=4$. (b) $\lambda_1=0.7, \lambda_2=0.8$, and $L_1=L_2=4$. (c) $\lambda_1=\lambda_2=0.7$, $L_1=4$, and $L_2=2$.}\label{fig:two}
% \vspace{-0.2cm}
\end{figure}

\subsection{Structure of the Optimal Policy in Section~\ref{sec:optimal policy}}
Fig.~\ref{fig:single} shows the structure of the optimal policy for a single IoT device for different values of the number of remaining packets $D$ for the current in-transmission status update. This figure focuses on the optimal sampling action $v^*$.
We can observe that the decision to start sending a new status update (i.e., $v^*=2$) is threshold-based with respect to $A_d$, which verifies the result in Theorem~\ref{theorem:optimal}.
From Fig.~\ref{fig:single_D3}, we can see that the decision of continuing to send the current in-transmission update (i.e., $v^*=1$) is not threshold-based with respect to $A_r$ \textcolor{black}{and is not threshold-based with respect to $D$.}
% The reason is that, for a large $D$, continuing to send the current in-transmission update is not much different from starting to send a new update in reducing a small $A_r$.
The reason is that, for a large $D$, an already small AoI $A_r$ will not be significantly improved if the device decides to stop sending its current update and, instead, it transmits a new update.

% \begin{figure}[!t]
% \begin{minipage}[h]{.475\linewidth}
% \centering
%        % \includegraphics[height=3.7cm, width=3.7cm]{symStruc3q.eps} [height=4cm, width=4.2cm]
%        \includegraphics[scale=0.5]{single_IoT_optimal_D1.eps}
% \subcaption{}\label{fig:single_D1}
% \end{minipage}
% \begin{minipage}[h]{.475\linewidth}
% \centering
%         % \includegraphics[height=3.7cm, width=3.7cm]{symStruc2q.eps}
%         \includegraphics[scale=0.5]{single_IoT_optimal_D3.eps}
% \subcaption{}\label{fig:single_D3}
% \end{minipage}
% % \vspace{-0.2cm}
% \caption{Structure of optimal policy $\pi^*$ in the single IoT device case. $L=4$. $\lambda = 0.8$. (a) $D=1$. (b) $D=3$.}\label{fig:single}
% % \vspace{-0.4cm}
% \end{figure}

Fig.~\ref{fig:two} illustrates the structure of the optimal policy for two IoT devices under different values for the channel reliability $\lambda_k$, for different packet sizes $L_k$. Here, we focus on the optimal scheduling action\textcolor{black}{\footnote{\textcolor{black}{We choose to schedule device 1 if scheduling device 1 achieves the same AoI performance with scheduling device 2.}}}.
It can be seen that, the scheduling action of different devices is of a switch-type structure.
Moreover, by comparing Fig.~\ref{fig:two_equal} with Fig.~\ref{fig:two_higher} and by comparing Fig.~\ref{fig:two_equal} with Fig.~\ref{fig:two_D_heter}, we can observe that the device having a better channel reliability or having a smaller packet size is given a higher scheduling priority.

\begin{figure}[!t]
\begin{minipage}[h]{0.475\linewidth}
\centering
       \includegraphics[scale=0.5]{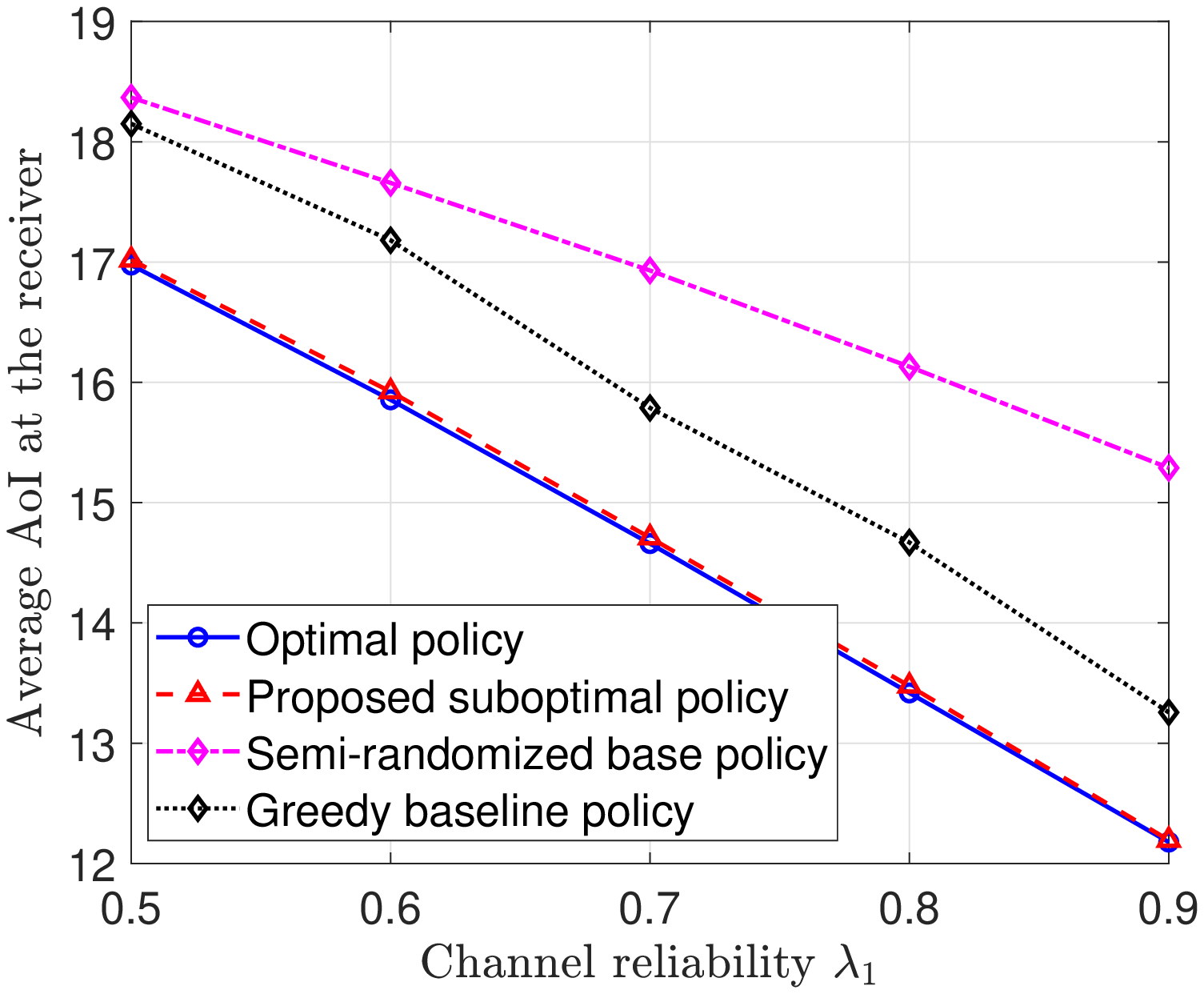}
\subcaption{}\label{fig:compare_two_uniform}
\end{minipage}
\begin{minipage}[h]{0.475\linewidth}
\centering
        \includegraphics[scale=0.5]{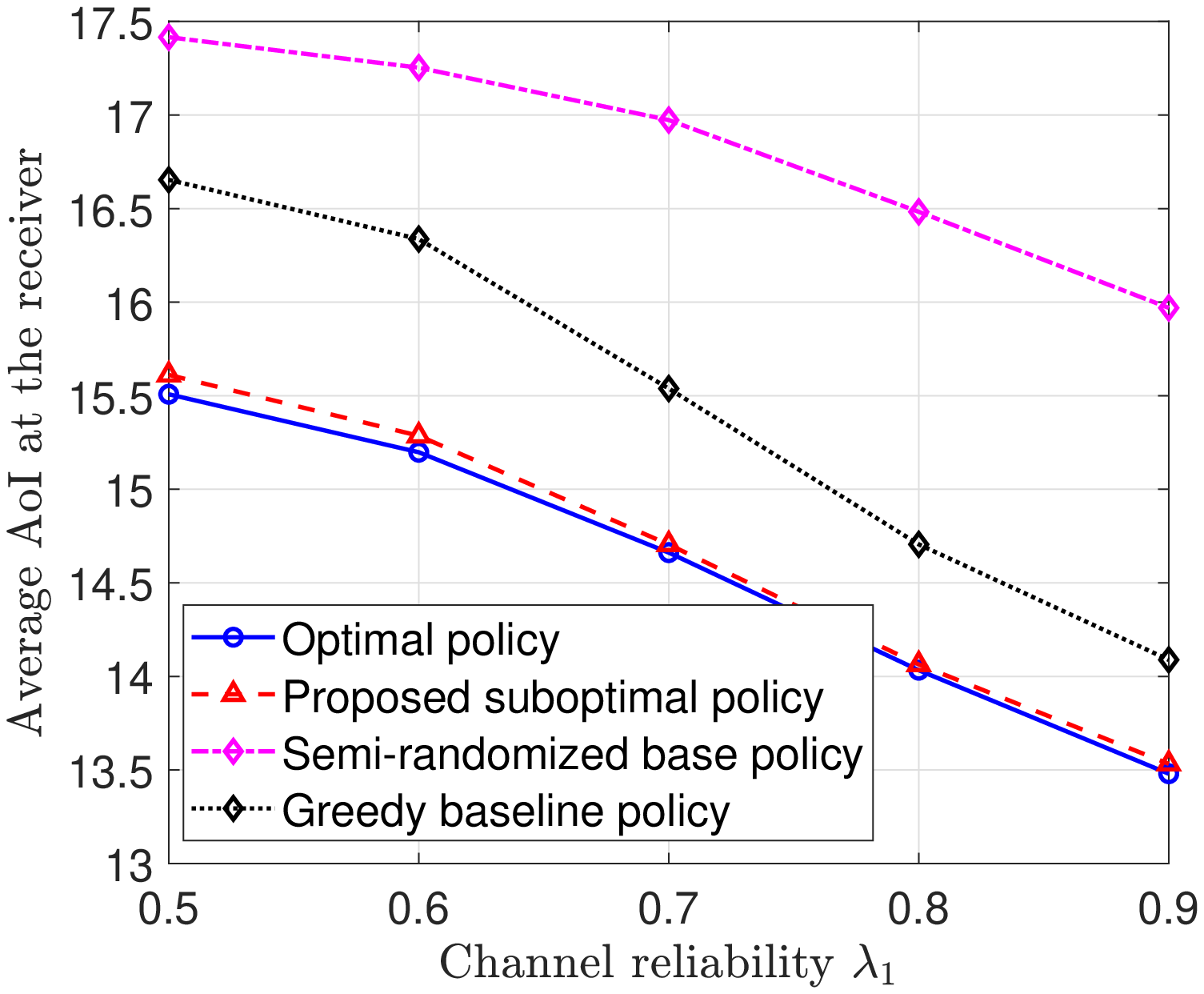}
\subcaption{}\label{fig:compare_two_nonuniform}
\end{minipage}
 % \vspace{-0.4cm}
\caption{Performance comparison among the optimal policy, the proposed suboptimal policy,  the semi-randomized base policy, \textcolor{black}{and the greedy baseline policy}.  $\hat{A}_{d,k}=\hat{A}_{r,k}=10$, for $k=1,2$. $L_1=L_2=3$. $M=1$. (a) $\lambda_1=\lambda_2$. (b) $\lambda_2=0.7$.}\label{fig:compare_two}
%$\hat{A}_{d,k}=\hat{A}_{r,k}=10$ for $k=1,2$.
 % \vspace{-0.5cm}
\end{figure}

\subsection{Performance of the Proposed Suboptimal Policy in Section~\ref{sec:suboptimal}}

In Fig.~\ref{fig:compare_two}, we compare the average \textcolor{black}{the AoI at the receiver}, resulting from the optimal policy $\pi^*$, the proposed suboptimal policy $\hat{\pi}^*$, the semi-randomized base policy $\hat{\pi}$, \textcolor{black}{and the greedy baseline policy} for two IoT devices under different channel reliability parameters.
% It can be observed that the average \textcolor{black}{the AoI at the receiver} achieved by the optimal policy and the proposed suboptimal policy are very close to each other, and is much lower than that of the semi-randomized base policy.
Fig.~\ref{fig:compare_two} shows that the proposed suboptimal policy achieves a near-optimal performance and significantly outperforms the semi-randomized base policy \textcolor{black}{and the greedy baseline policy}.
This stems from the structural similarity between the proposed suboptimal policy and the optimal policy. Hence,  the proposed suboptimal policy can make foresighted decision by better exploiting the system state information and channel statistics.

\begin{figure}[!t]
\begin{minipage}[h]{0.475\linewidth}
\centering
       \includegraphics[scale=0.5]{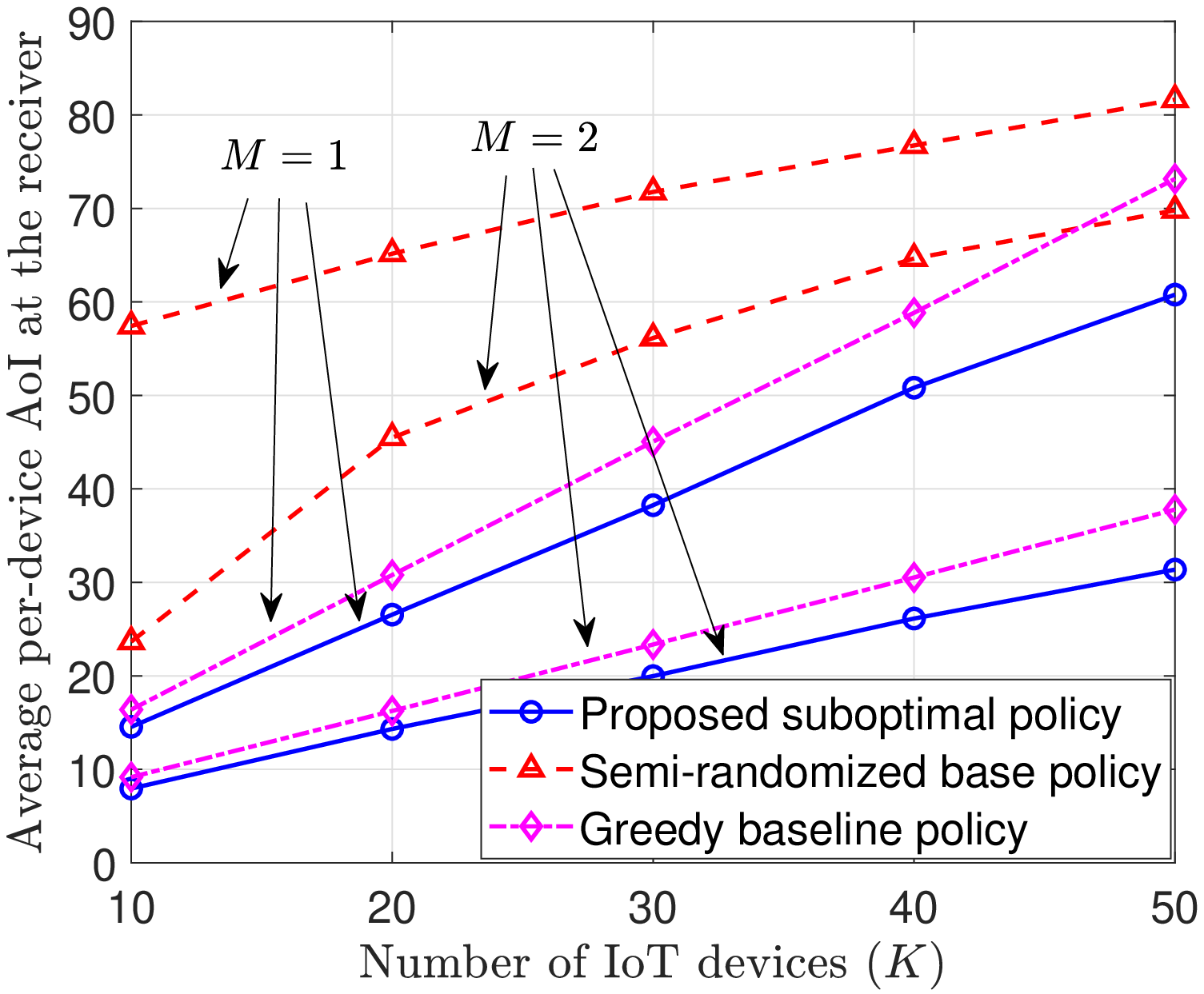}
\subcaption{}\label{fig:compare_vs_K_uni}
\end{minipage}
\begin{minipage}[h]{0.475\linewidth}
\centering
        \includegraphics[scale=0.5]{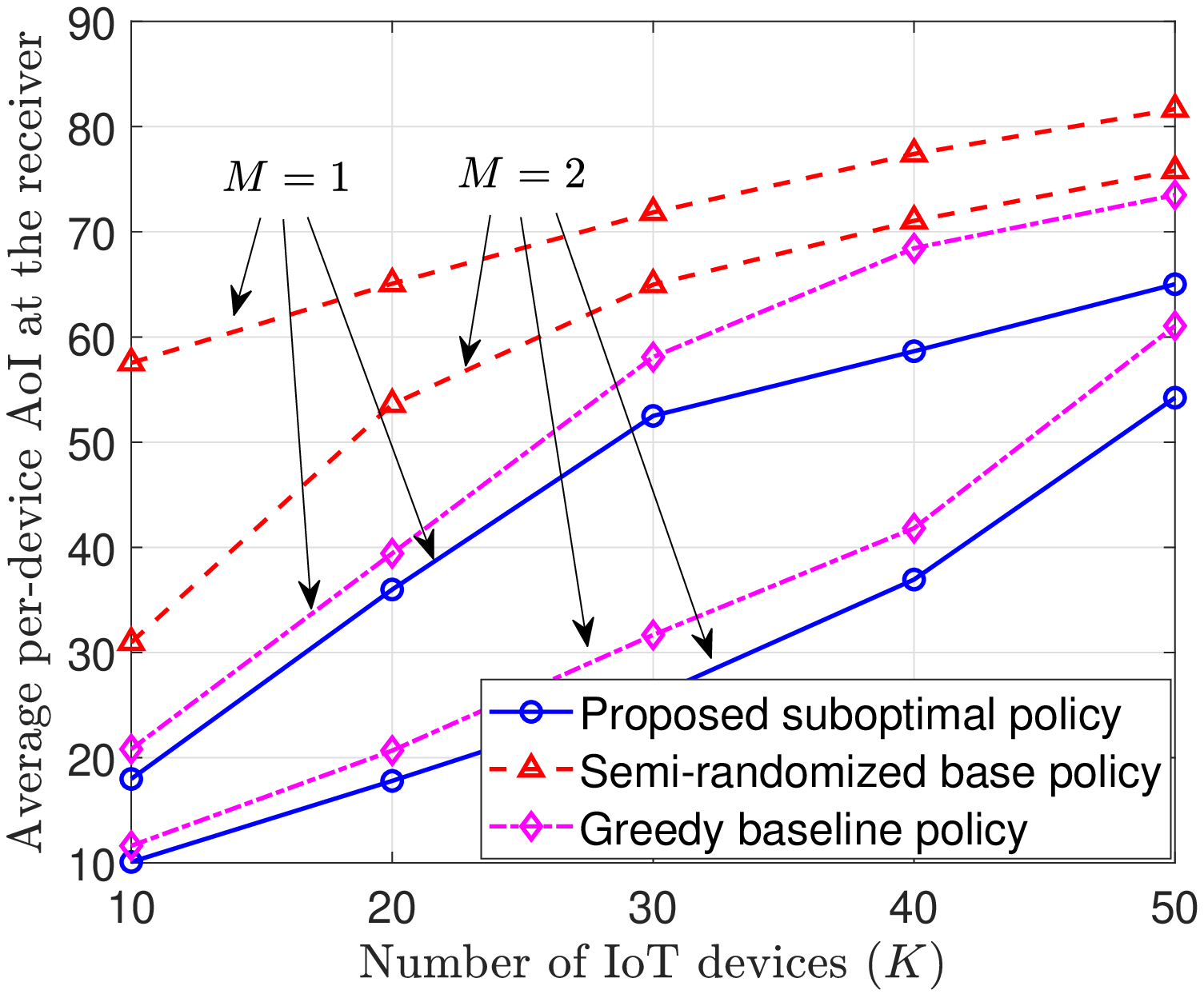}
\subcaption{}\label{fig:compare_vs_K_nonuni}
\end{minipage}
 % \vspace{-0.4cm}
\caption{Average per-device AoI at the proposed suboptimal policy, the semi-randomized base policy, \textcolor{black}{and the greedy baseline policy} versus the number of IoT devices $K$. $\hat{A}_{d,k}=\hat{A}_{r,k}=100$ and $\lambda_k=0.8$, for all $k$. (a) Uniform case. (b) Nonuniform case. }\label{fig:compare_vs_K}
%$\hat{A}_{d,k}=\hat{A}_{r,k}=10$ for $k=1,2$.
 % \vspace{-0.5cm}
\end{figure}

\begin{figure}[!t]
\begin{minipage}[h]{0.475\linewidth}
\centering
       \includegraphics[scale=0.5]{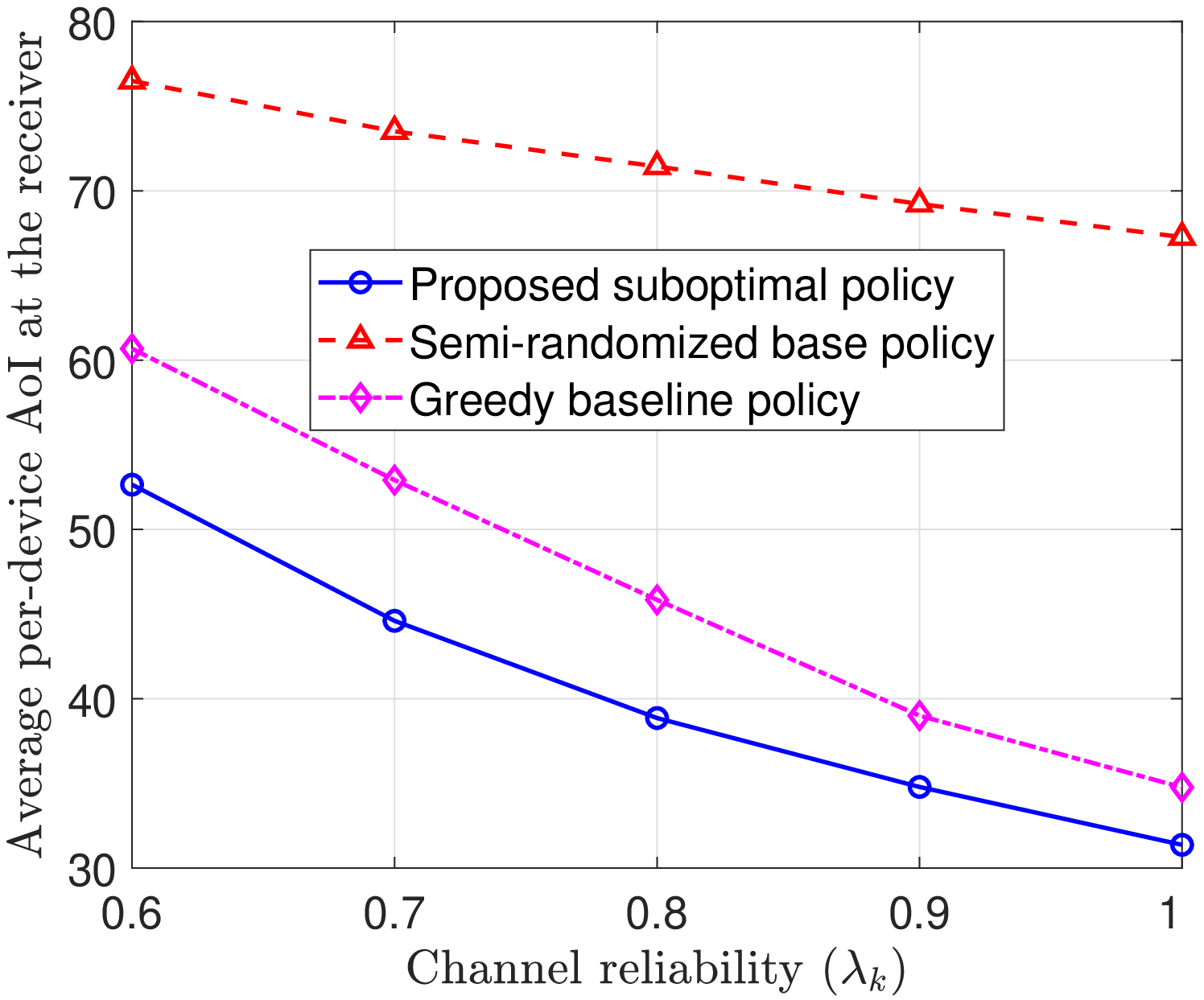}
\subcaption{}\label{fig:compare_vs_lambda_uni}
\end{minipage}
\begin{minipage}[h]{0.475\linewidth}
\centering
        \includegraphics[scale=0.5]{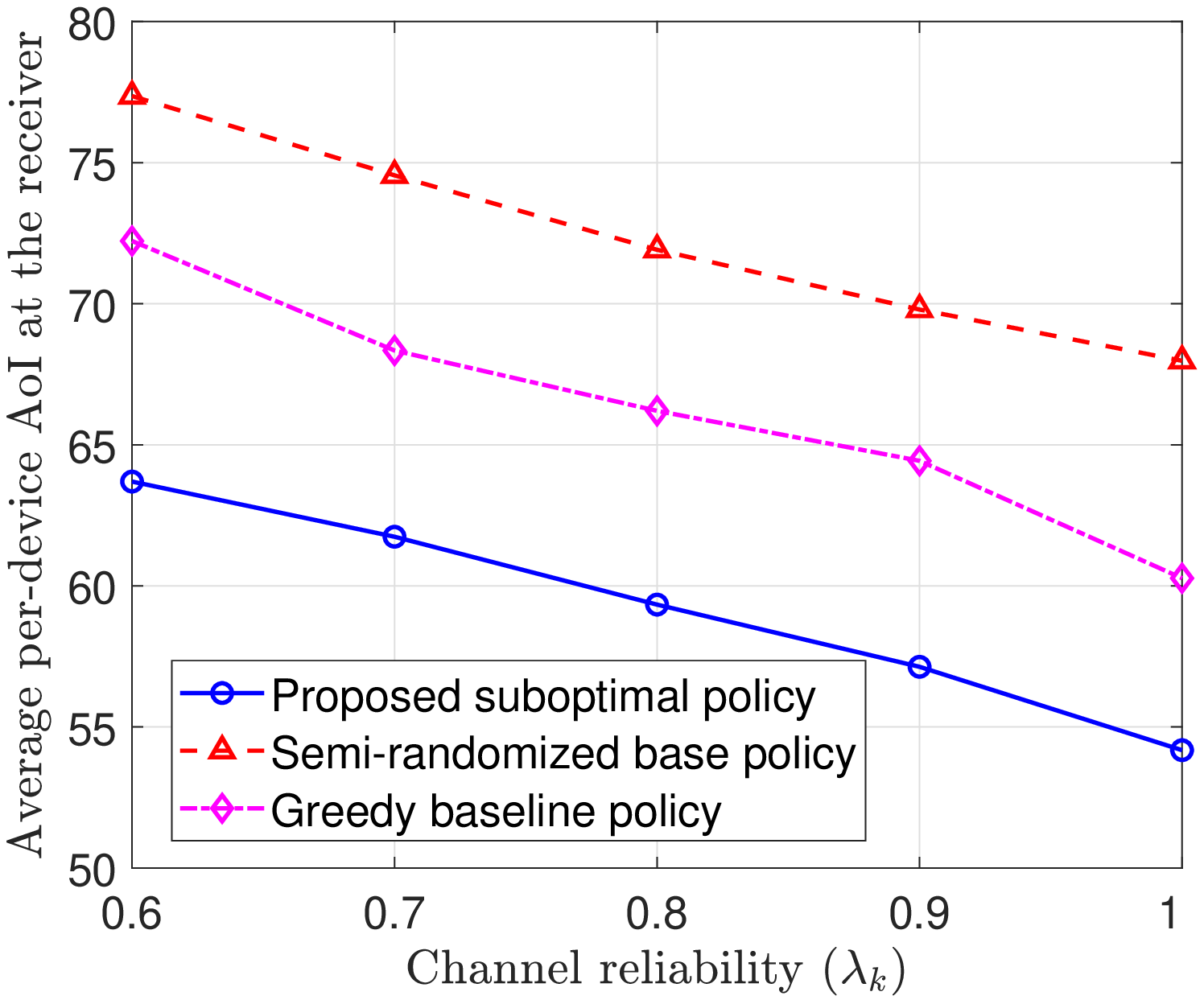}
\subcaption{}\label{fig:compare_vs_lambda_nonuni}
\end{minipage}
 % \vspace{-0.4cm}
\caption{Average per-device AoI at the proposed suboptimal policy, the semi-randomized base policy, \textcolor{black}{and the greedy baseline policy} versus the channel reliability $\lambda_k$.  $K=30$. $\hat{A}_{d,k}=\hat{A}_{r,k}=100$, for all $k$, $\lambda_1=\lambda_2=\cdots=\lambda_K$. (a) Uniform case. (b) Nonuniform case. }\label{fig:compare_vs_lambda}
%$\hat{A}_{d,k}=\hat{A}_{r,k}=10$ for $k=1,2$.
 % \vspace{-0.5cm}
\end{figure}

Next, we investigate the effects of varying the number of IoT devices $K$, the number of maximum allowed scheduled IoT devices $M$, and the channel reliability of IoT devices $\lambda_k$, on the AoI performance of the proposed suboptimal policy and the semi-randomized base policy.  
\textcolor{black}{Note that the computational complexity needed to obtain the optimal policy is prohibitively high for large values of $K$, $\hat{A}_{d,k}$ and  $\hat{A}_{r,k}$, due to the curse of dimensionality and, thus, we could not derive the optimal policy for these cases.}
% Note that, for the system with many IoT devices, we cannot derive the optimal policy due to its prohibitively high computational cost.% Hereinafter, we set the $\hat{A}_{d,k}=\hat{A}_{r,k}=100$ for all $k$.
 The simulation results are obtained by averaging over 10,000 time slots. We consider the uniform and nonuniform cases, based on whether the packet sizes for the IoT devices $L_k$ are the same or not. Particularly, for the uniform case, we set $\lambda_k=2$ for all $k$, and for the nonuniform case, we set $\lambda_k=2$ for $k=1,\cdots,K/2$ and $\lambda_k=3$ for $k=K/2+1,\cdots,K$.

Fig.~\ref{fig:compare_vs_K} illustrates the average, per-device \textcolor{black}{the AoI at the receiver} resulting from the proposed suboptimal policy, the semi-randomized base policy, \textcolor{black}{and the greedy baseline policy,} for different numbers of IoT devices $K$ and maximum allowed scheduled IoT devices $M$.
From Fig.~\ref{fig:compare_vs_K_uni}, we can see that the proposed suboptimal policy can reduce the average \textcolor{black}{the AoI at the receiver} by up to 74\% \textcolor{black}{and 17\%,} compared to the semi-randomized base policy \textcolor{black}{and the greedy baseline policy, respectively, for $M=1$.}
Moreover, for \textcolor{black}{all} policies, the average per-device \textcolor{black}{the AoI at the receiver} increases when $K$ increases and decreases when $M$ increases.
This is because the transmission opportunities for each IoT device decrease with $K$ and increase with $M$.

\begin{figure}[!t]
\begin{minipage}[h]{.475\linewidth}
\centering
       \includegraphics[scale=0.49]{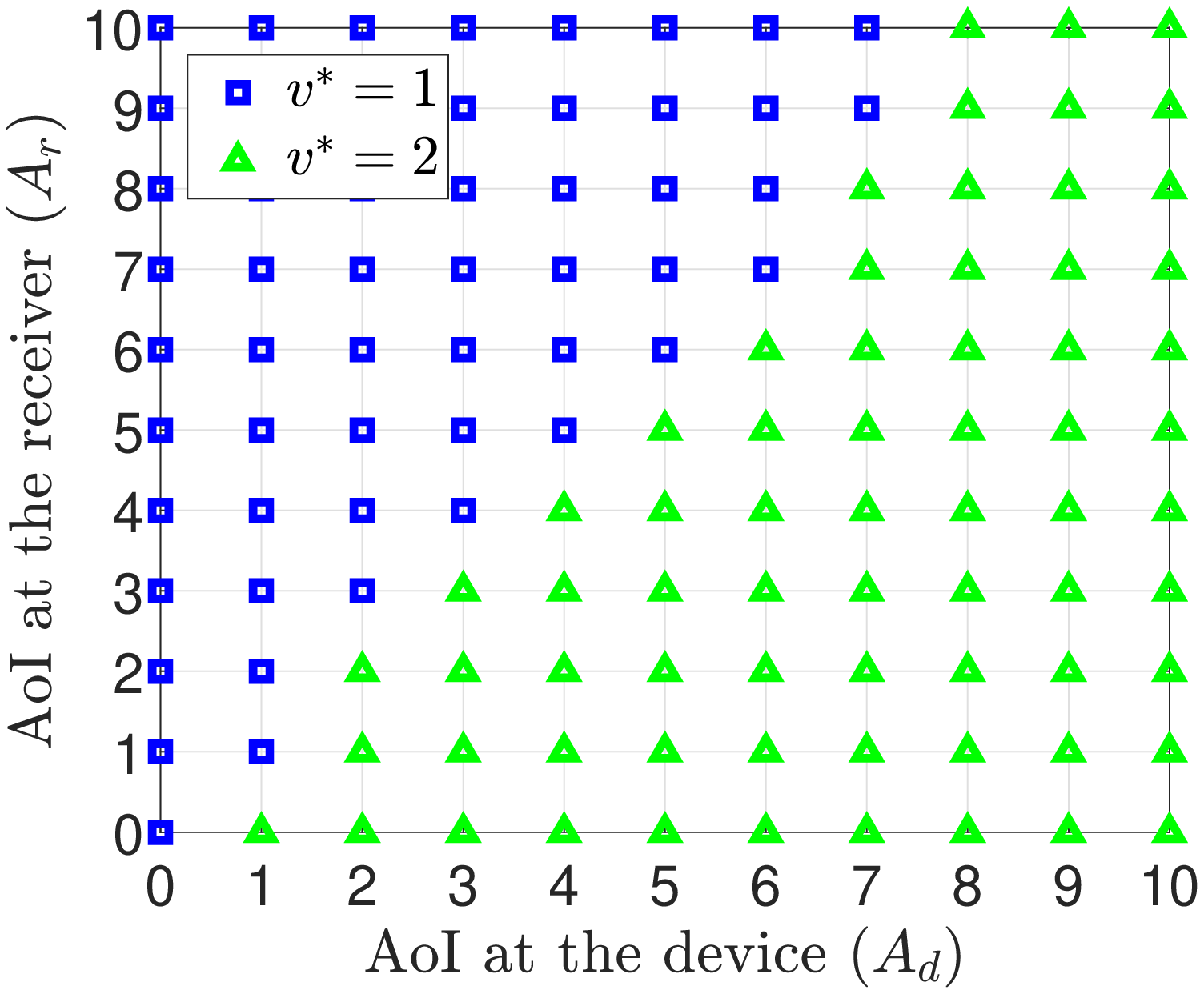}
\subcaption{}\label{fig:single_D1rho5}
\end{minipage}
\begin{minipage}[h]{.475\linewidth}
\centering
        \includegraphics[scale=0.49]{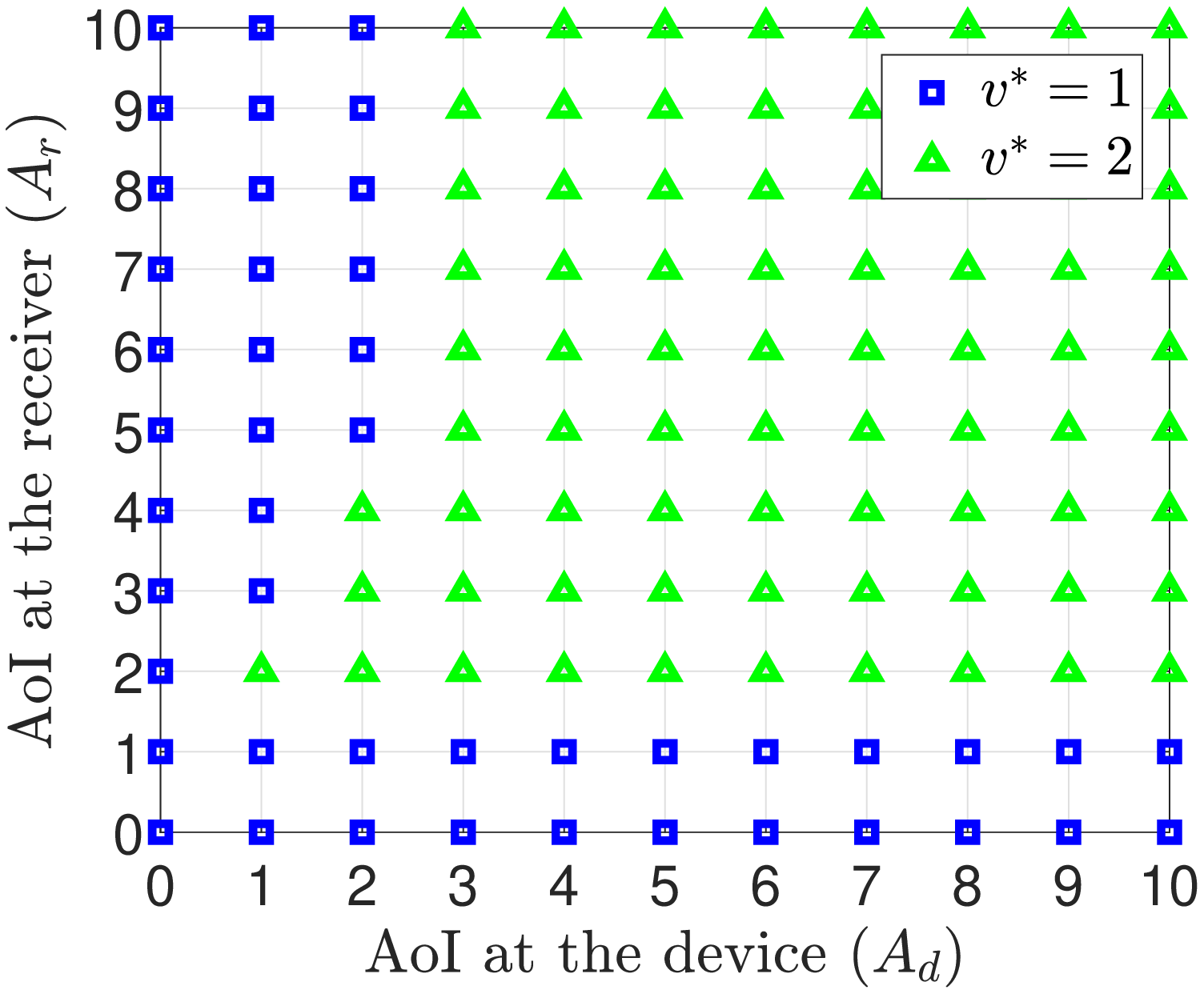}
\subcaption{}\label{fig:single_D3rho5}
\end{minipage}
\begin{minipage}[h]{\linewidth}
\centering
        \includegraphics[scale=0.49]{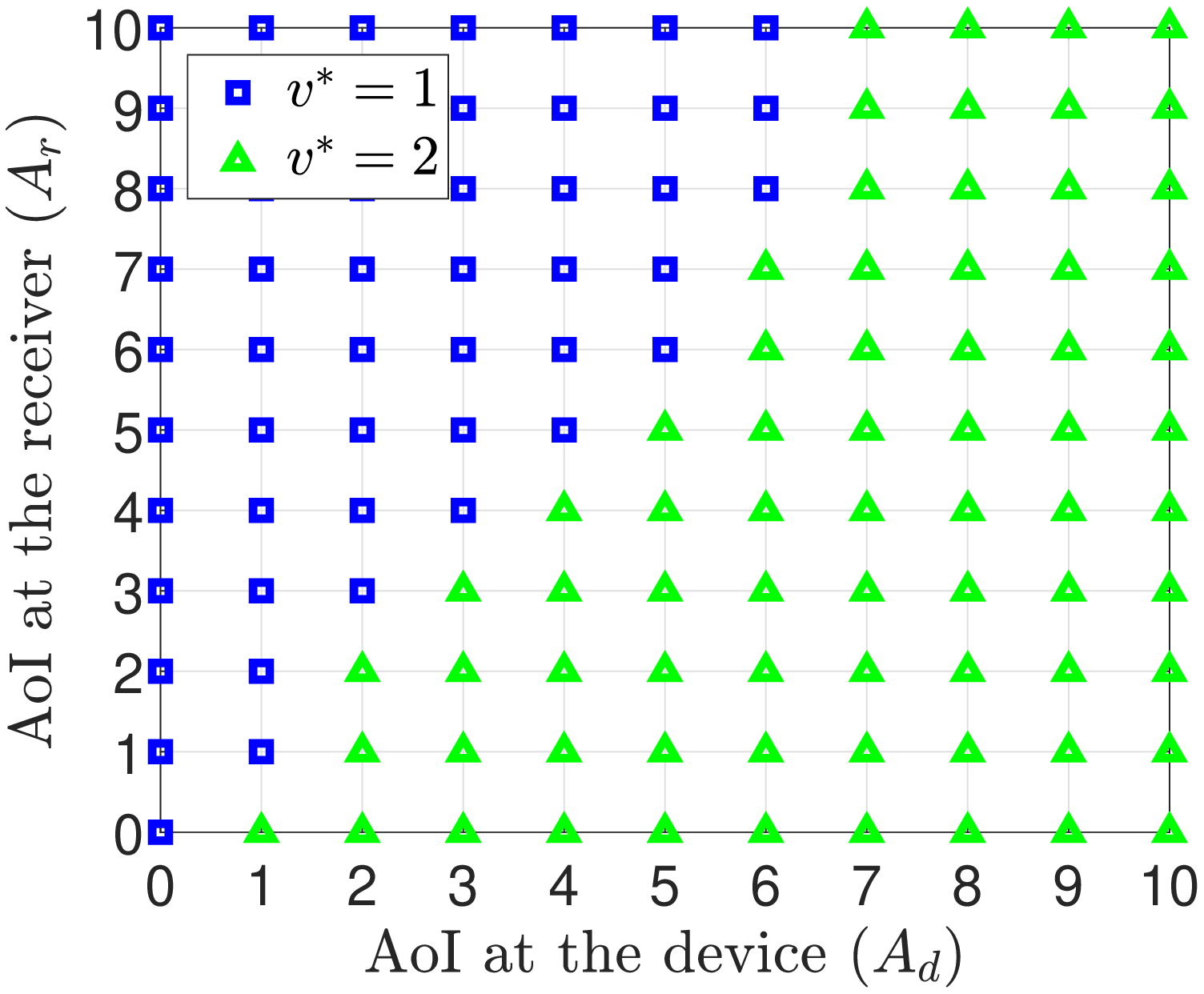}
\subcaption{}\label{fig:single_D1rho8}
\end{minipage}
% \vspace{-0.4cm}
\caption{Structure of optimal policy $\pi^*$ in the single IoT device case with random status update arrivals. $\hat{A}_b=\hat{A}_l=\hat{A}_r=10$, $L=4$, $A_b=3$, and $\lambda=0.8$. (a) $D=1$ and $\rho=0.5$. (b) $D=3$ and $\rho=0.5$. (c) $D=1$ and $\rho=0.8$.}\label{fig:single_extension}
% \vspace{-0.3cm}
\end{figure}

Fig.~\ref{fig:compare_vs_lambda} shows the average, per-device \textcolor{black}{the AoI at the receiver} resulting from the proposed suboptimal policy, the semi-randomized base policy, \textcolor{black}{and the greedy baseline policy}, under different channel reliability of IoT devices $\lambda_k$.
From Fig.~\ref{fig:compare_vs_lambda_uni}, we observe that the average AoI reduction achieved by the proposed suboptimal policy compared to the semi-randomized base policy \textcolor{black}{and the greedy baseline policy} can be as much as 53\% \textcolor{black}{and 16\%, respectively}.
Moreover, Fig.~\ref{fig:compare_vs_lambda} shows that, when  $\lambda_k$ increases,  the average \textcolor{black}{the AoI at the receiver} for \textcolor{black}{all} policies will decrease.
% This indicates that the AoI performance improves with the wireless channel condition.
This is intuitive as  channels with better quality, i.e., larger $\lambda_k$, will achieve a smaller \textcolor{black}{the AoI at the receiver}.

\subsection{Structure of the Optimal Policy in Section~\ref{sec:extension}}
In Fig.~\ref{fig:single_extension}, we illustrate the structure of the optimal policy for the IoT system with random status update arrivals in a single IoT device case. We also focus on the optimal sampling action $v^*$.
 From Fig.~\ref{fig:single_extension}, we can see that the decision to start sending a new status update (i.e., $v^*=2$) is threshold-based with respect to $A_d$.
 This verifies the result of Theorem~\ref{theorem:optimal_extension}.
By comparing Fig.~\ref{fig:single_D1rho5} with Fig.~\ref{fig:single_D1rho8}, we further observe that, the IoT device is more likely to start transmitting the status update in the buffer, when the arrival rate of the status updates is larger.
This is because the the status update in the buffer will be refreshed more frequently for a large arrival rate of the status updates, and, thus, could be more beneficial to the destination.

\section{Conclusion}
In this paper, we have studied the problem of optimal device scheduling and status sampling policy that minimizes the average AoI  for a real-time IoT monitoring system with non-uniform sizes of status updates and noisy channels.
We have formulated this problem as an infinite horizon average cost MDP.
By characterizing the monotonicity property of the value function, we have shown that the optimal policy is threshold-based with respect to the AoI at each IoT device.
To reduce the complexity in computing the optimal policy, we have proposed a low-complexity suboptimal policy based on a semi-randomized base policy and linear approximated value functions.
We have shown that the proposed suboptimal policy has a similar threshold structure to the optimal policy, which serves as a structural base for its good performance.
Then, we have extended those analytical results to the IoT monitoring system, where the status updates cannot be generated at will by the IoT devices and can only randomly arrive at the devices.

Simulation results have shown that, for the IoT system without random status update arrivals,  the optimal policy is not threshold-based with respect to \textcolor{black}{the AoI at the receiver} for each device, the device with better channel reliability can have a higher scheduling priority, and  the proposed suboptimal policy can achieve a near-optimal AoI performance and significantly outperforms the semi-randomized base policy.
Moreover, the results have shown that, for the IoT system with random status update arrivals,  the device is more willing to start sending the status update in the buffer for a large arrival rate of status updates.
\textcolor{black}{Future work can extend the developed algorithms to scenarios in which each device has explicit energy limitations.}

\appendices
% \section*{Appendix}

% \section{Proof of Lemma~\ref{lemma:bellman}}\label{app:bellman}
\begin{answer}
\section{Extension to Non-zero Generation Time of Status Updates}\label{app:non-zero generation time}
% We extend the framework to the scenario in which the generation time of status updates is non-zero.
Let $\tau_k\geq 1$ be the number of time slots needed to generate a status update at device $k$. 
Note that, if  device $k$ is scheduled to sample at slot $t$, then it needs to wait at least $\tau_k$ slots before starting to transmit this newly generated status update.
Here, for each device $k$, we let $D_k(t)\in\{1,\cdots,L_k,\cdots,L_k+\tau_k-1\}$, where $D_k(t)$ denotes the number of the remaining status packets if $D_k(t)\leq L_k$ and $D_k(t)$ denotes the minimum number of slots that device $k$ needs so as to deliver the status update if $D_k(t)>L_k$. 
Note that, if $D_k(t)>L_k$, there will be no  status update that is available for transmission, i.e., $w_k(t)\neq (1,1)$.

We need to consider the following four possible cases: a) When device $k$ is scheduled to continue the current in-transmission status update at slot $t$ (i.e., $\bs{w}(t)=(1,1)$ and $D_k(t)\leq L_k$) and the transmission succeeds at slot $t$, then, if there is only one remaining packet at $t$ (i.e., $D_k(t)=1$), $D_k(t+1)$ will be reset to $L_k+\tau_k-1$; otherwise, $D_k(t+1)$ will be $D_k(t)-1$.
b) When device $k$ is scheduled to continue the current in-transmission status update at slot $t$ (i.e., $\bs{w}(t)=(1,1)$ and $D_k(t)\leq L_k$) and the transmission fails at slot $t$, $D(t+1)$ will still be $D_k(t)$.
c) When device $k$ is scheduled to sample at slot $t$ (i.e., $\bs{w}(t)=(1,2)$), then $D(t+1)$ will be $L_k+\tau_k-1$.
d) When device $k$ is not scheduled (i.e., $\bs{w}(t)=(0,0)$), then, if $D_k(t)>L_k$, then $D_k(t+1)$ will be $D_k(t)-1$, otherwise, $D(t+1)$ will still be $D_k(t)$. In summary, for each device $k$, we can now define the dynamics of $D_k(t)$ as follows:
\begin{align}\label{eqn:d_device_k_extension}
D_k(t+1)
&=\begin{cases} &\mathbbm{1}(D_k(t)=1)(L_k+\tau_k-1)+\mathbbm{1}(D_k(t)>1)(D_k(t)-1),  \\
&\hspace{29mm}~\text{if}~\bs{w}_k(t)=(1,1)~\text{and~transmission succeeds at}~t,\\
% &L_k-1, ~ \text{if}~ \bs{w}_k(t)=(1,2)~\text{and~transmission succeeds}\\
% &\hspace{70mm}\text{ at}~t,\\
&L_k+\tau_k-1, \hspace{7mm}~ \text{if}~\bs{w}_k(t)=(1,2),\\
&\mathbbm{1}(D_k(t)>L_k)(D_k(t)-1)+\mathbbm{1}(D_k(t)\leq L_k)D_k(t), \hspace{5mm}~ \text{otherwise.}
      \end{cases}
\end{align}

For the AoI at device $k$, when device $k$ is scheduled to continue the current in-transmission status update at slot $t$ (i.e., $\bs{w}(t)=(1,1)$ and $D_k(t)\leq L_k$), or device $k$ is scheduled to sample at slot $t$, the AoI will decrease to zero; otherwise, the AoI will increase by one. Thus, the dynamics of the AoI at device $k$ will be given by:
\begin{align}\label{eqn:aoi_device_k_extension}
A_{d,k}(t+1)
&=\begin{cases} &0,  ~\text{if}~\bs{w}_k(t)=(1,1), D_k(t) = 1, \text{and~transmission succeeds at}~t;\\
&\hspace{8mm}~\text{or}~\bs{w}_k(t)=(1,2),\\
                &\min\{A_{d,k}(t)+1,\hat{A}_{d,k}\}, ~ \text{otherwise.}
      \end{cases}
\end{align}
The dynamics of the destination's AoI $A_{r,k}(t)$ of device $k$ are the same in \eqref{eqn:aoi_bs_k}.

It is obvious that, if $D_k(t)>L_k$, there is no status update for device $k$ to send (i.e., $\bs{w}_k(t)\neq(1,1)$) and there is no need to re-sample another new status update during the generation of the previous status update (i.e., $\bs{w}_k(t)\neq(1,2)$). Thus, we set $\bs{w}_k(t)=(0,0)$ if  $D_k(t)>L_k$.

Then, we can formulate the MDP in the same manner in Section~\ref{sec:problem formulation}.
It can be easily verified that the monotonicity of the value function still holds and we can obtain the exact same structural properties of the optimal policy to the one in Theorem~\ref{theorem:optimal},  by following the line of the analysis in Section~\ref{sec:optimal policy}.
The suboptimal solution in Section~\ref{sec:suboptimal} can also be readily extended for non-zero generation time.
\end{answer}

\section{Proof of Lemma~\ref{lemma:propertyV}}\label{app:propertyV}

We prove Lemma~\ref{lemma:propertyV} using the relative value iteration algorithm (RVIA) \textcolor{black}{\cite[Chapter 5.3]{bertsekas4}} and mathematical induction.
First, we present the RVIA. For each system state $\bs{X}\in\mathcal{X}$, we denote by $V_n(\bs{X})$ the value function at iteration $n$, where $n=1,2,\cdots$. Define the state-action cost function at iteration $n$ as:
\begin{align}
  J_n(\bs{X},\bs{w})=\sum_{k=1}^KA_{r,k} + \sum_{\bs{X}'\in\mathcal{X}}\Pr[\bs{X}'|\bs{X},\bs{w}] V_n(\bs{X}').\label{eqn:J_function_n}
\end{align}
where $\Pr[\bs{X}'|\bs{X},\bs{w}]$ is given by \eqref{eqn:trans_prob}. Note that $J_n(\bs{X},\bs{w})$ is related to the right-hand side of the Bellman equation in \eqref{eqn:bellman}.
For each $\bs{X}$, RVIA can be used to find $V_n(\bs{X})$ according to:
\begin{equation}\label{eqn:RVIA}
  V_{n+1}(\bs{X})=\min_{\bs{w}\in\mathcal{W}} J_{n+1}(\bs{X},\bs{w})- \min_{\bs{w}\in\mathcal{W}}J_{n+1}(\bs{X}^{\dag},\bs{w}),~\forall n,
\end{equation}
where $\bs{X}^{\dag}$ is some fixed state. According to \textcolor{black}{\cite[Proposition 5.3.2]{bertsekas4}}, the generated sequence $\{V_n(\bs{X})\}$ converges to $\{V(\bs{X})\}$, under any initialization of $V_0(\bs{X})$, i.e.,
 \begin{equation}
   \lim_{n\to\infty}V_n(\bs{X})=V(\bs{X}),~\forall \bs{X}\in\mathcal{X},\label{eqn:converge}
 \end{equation}
where $V(\bs{X})$ satisfies the Bellman equation in \eqref{eqn:bellman}.
Let $\pi^*_n(\bs{X})$ be the control action attains the minimum of the first term in \eqref{eqn:RVIA} at iteration $n$ for all $\bs{X}$, i.e.,
\begin{equation}
  \pi^*_n(\bs{X})=\arg\min_{\bs{w}\in\mathcal{W}} J_{n+1}(\bs{X},\bs{w}),~~\forall\bs{X}\in\mathcal{X}.\label{eqn:optimal_n}
\end{equation}
Define $\pi^*_n(\bs{X})\triangleq (\pi^*_{n,k}(\bs{X}))_{k\in\mathcal{K}}$, where $\pi^*_{n,k}(\bs{X})$ denotes the control action of IoT device $k$ under state $\bs{X}$.
We refer to $\pi^*_n$ as the optimal policy at iteration $n$.

Now, we prove Lemma~\ref{lemma:propertyV} through the RVIA using mathematical induction.
Consider two system states $\bs{X}^1=(\bs{A}_d^1,\bs{A}_r^1,\bs{D}^1)$ and $\bs{X}^2=(\bs{A}_d^2,\bs{A}_r^2,\bs{D}^2)$. To prove Lemma~\ref{lemma:propertyV}, according to \eqref{eqn:converge}, it suffices to show that for any $\bs{X}^1$ and $\bs{X}^2$ such that $\bs{A}_d^2\succeq\bs{A}_d^1$, $\bs{A}_r^2\succeq\bs{A}_r^1$, and $\bs{D}^2=\bs{D}^1$,
\begin{equation}
V_{n}(\bs{X}^2)\geq V_{n}(\bs{X}^1),\label{eqn:vn}
\end{equation}
holds for all $n=1,2,\cdots$.

First, we initialize $V_1(\bs{X})$ for all $\bs{X}$. Thus, \eqref{eqn:vn} holds for $n=1$.
Assume \eqref{eqn:vn} holds for some $n>1$. We will show that \eqref{eqn:vn} holds for $n+1$. By \eqref{eqn:RVIA}, we have
\begin{align}
V_{n+1}(\bs{X}^1) &= J_{n+1}(\bs{X}^1,\pi^*_n(\bs{X}^1)) - J_{n+1}(\bs{X}^{\dag},\pi^*_n(\bs{X}^{\dag}))\nonumber\\
&\overset{(a)}{\leq} J_{n+1}(\bs{X}^1,\pi^*_n(\bs{X}^2)) - J_{n+1}(\bs{X}^{\dag},\pi^*_n(\bs{X}^{\dag}))\nonumber\\
&=  \sum_{k} A_{d,k}^1 + \sum_{\bs{X}^{1'}\in\mathcal{X}}\Pr[\bs{X}^{1'}|\bs{X}^1,\pi^*_n(\bs{X}^2)] V(\bs{X}^{1'}) - J_{n+1}(\bs{X}^{\dag},\pi^*_n(\bs{X}^{\dag})),
\end{align}
where $(a)$ is due to the optimality of $\pi^*_n(\bs{X}^1)$ for $\bs{X}^1$ at iteration $n$. %and $(b)$ directly follows from \eqref{eqn:J_function_n}.
By \eqref{eqn:J_function_n} and \eqref{eqn:RVIA},  we have
\begin{align}
V_{n+1}(\bs{X}^2) &= J_{n+1}(\bs{X}^2,\pi^*_n(\bs{X}^2)) - J_{n+1}(\bs{X}^{\dag},\pi^*_n(\bs{X}^{\dag}))\nonumber\\
&= \sum_{k} A_{d,k}^2 + \sum_{\bs{X}^{2'}\in\mathcal{X}}\Pr[\bs{X}^{2'}|\bs{X}^2,\pi^*_n(\bs{X}^2)] V(\bs{X}^{2'}) - J_{n+1}(\bs{X}^{\dag},\pi^*_n(\bs{X}^{\dag})).
\end{align}

We compare $\sum_{\bs{X}^{1'}\in\mathcal{X}}\Pr[\bs{X}^{1'}|\bs{X}^1,\pi^*_n(\bs{X}^2)] V(\bs{X}^{1'})$ with $\sum_{\bs{X}^{2'}\in\mathcal{X}}\Pr[\bs{X}^{2'}|\bs{X}^2,\pi^*_n(\bs{X}^2)] V(\bs{X}^{2'})$ for all possible $\pi^*_n(\bs{X}^2)=(\pi^*_{n,k}(\bs{X}^2))_{k\in\mathcal{K}}$.
For each $k$, we need to consider the following three cases for $\pi^*_{n,k}(\bs{X}^2)$, i.e., $\pi^*_{n,k}(\bs{X}^2)=(0,0), (1,1), (1,2)$. According to \eqref{eqn:trans_prob_k}, we can check that $X_{d,k}^{2'}\geq X_{d,k}^{1'}$, $X_{r,k}^{2'}\geq X_{r,k}^{1'}$, and $D_{k}^{2'}= D_{k}^{1'}$ hold for each of the three cases.  Thus, by the induction hypothesis, we have $\sum_{\bs{X}^{2'}\in\mathcal{X}}\Pr[\bs{X}^{2'}|\bs{X}^2,\pi^*_n(\bs{X}^2)] V(\bs{X}^{2'})\geq \sum_{\bs{X}^{1'}\in\mathcal{X}}\Pr[\bs{X}^{1'}|\bs{X}^1,\pi^*_n(\bs{X}^2)] V(\bs{X}^{1'})$, which implies that $V_{n+1}(\bs{X}^2)\geq V_{n+1}(\bs{X}^1)$, i.e., \eqref{eqn:vn} holds for $n+1$.
Therefore, by induction, we know that \eqref{eqn:vn} holds for any $n$. By taking limits on both sides of \eqref{eqn:vn} and by \eqref{eqn:converge}, we complete the proof of Lemma~\ref{lemma:propertyV}.

% If $\pi^*_{n,k}(\bs{X}^2)=(0,0)$, i.e., device $k$ is not scheduled, then, according to \eqref{eqn:trans_prob_k}, we have $\bs{X}^1_{k,un}=(\min\{A_{d,k}^1+1,\hat{A}_{d,k}\},\min\{A_{r,k}^1+1,\hat{A}_{r,k}\}, D_k^1)$ and $\bs{X}^2_{k,un}=(\min\{A_{d,k}^2+1,\hat{A}_{d,k}\},\min\{A_{r,k}^2+1,\hat{A}_{r,k}\}, D_k^2)$.
% If $\pi^*_{n,k}(\bs{X}^2)=(1,1)$,  we can obtain that, when $D_k^1=D_k^2=1$, then $\bs{X}_{k,s}^1=(0,\min\{A_{d,k}^1+1,\hat{A}_{r,k}\}, L_k)$ and $\bs{X}_{k,s}^2=(0,\min\{A_{d,k}^2+1,\hat{A}_{r,k}\}, L_k)$; when $D_k^1=D_k^2>1$, then $\bs{X}_{k,s}^1=\min\{A_{d,k}^1+1,\hat{A}_{d,k}\},\min\{A_{r,k}^1+1,\hat{A}_{r,k}\}, D_k^1-1$ and $\bs{X}_{k,s}^2=\min\{A_{d,k}^2+1,\hat{A}_{d,k}\},\min\{A_{r,k}^2+1,\hat{A}_{r,k}\}, D_k^2-1$, and $\bs{X}^1_{k,f}=(\min\{A_{d,k}^1+1,\hat{A}_{d,k}\},\min\{A_{r,k}^1+1,\hat{A}_{r,k}\}, D_k^1)$ and $\bs{X}^2_{k,f}=(\min\{A_{d,k}^2+1,\hat{A}_{d,k}\},\min\{A_{r,k}^2+1,\hat{A}_{r,k}\}, D_k^2)$

\section{Proof of Theorem~\ref{theorem:optimal}}\label{app:optimal}
To prove Theorem~\ref{theorem:optimal}, we first show that, for any $\bs{X}^1,\bs{X}^2\in\mathcal{X}$ and $\bs{w}\in\mathcal{W}$ such that $\bs{A}_r^1=\bs{A}_r^2$, $\bs{D}^1=\bs{D}^2$, and
\begin{equation}
\begin{cases}A_{d,k}^1\geq A_{d,k}^2,  & \text{if}~\bs{w}_k=(1,2), \\
            A_{d,k}^1=A_{d,k}^2,  &\text{otherwise},
  \end{cases},
\end{equation}
for all $k\in\mathcal{K}$,
\begin{align}\label{eqn:J_property}
J(\bs{X}^1,\bs{w}) - J(\bs{X}^1,\bs{w}') \leq J(\bs{X}^2,\bs{w}) - J(\bs{X}^2,\bs{w}')
\end{align}
holds for all $\bs{w}'\in\mathcal{W}$ and $\bs{w}'\neq \bs{w}$.
By \eqref{eqn:J_function}, we have
\begin{align}
&J(\bs{X}^1,\bs{w}) - J(\bs{X}^1,\bs{w}') - (J(\bs{X}^2,\bs{w}) - J(\bs{X}^2,\bs{w}'))\nonumber\\
 =&\underbrace{\sum_{\bs{X}^{1,\bs{w}}\in\mathcal{X}}\Pr[\bs{X}^{1,\bs{w}}|\bs{X}^1,\bs{w}] V(\bs{X}^{1,\bs{w}})}_{A} - \underbrace{\sum_{\bs{X}^{1,\bs{w}'}\in\mathcal{X}}\Pr[\bs{X}^{1,\bs{w}'}|\bs{X}^1,\bs{w}'] V(\bs{X}^{1,\bs{w}'})}_{B}\nonumber\\
 &-\underbrace{\sum_{\bs{X}^{2,\bs{w}}\in\mathcal{X}}\Pr[\bs{X}^{2,\bs{w}}|\bs{X}^2,\bs{w}] V(\bs{X}^{2,\bs{w}})}_{C}+ \underbrace{\sum_{\bs{X}^{2,\bs{w}'}\in\mathcal{X}}\Pr[\bs{X}^{2,\bs{w}'}|\bs{X}^2,\bs{w}'] V(\bs{X}^{2,\bs{w}'})}_{D}.
\end{align}
% & = \sum_{\bs{X}^{1,\bs{w}}\in\mathcal{X}}\Pr[\bs{X}^{1,\bs{w}}|\bs{X}^1,\bs{w}] V(\bs{X}^{1,\bs{w}}) - \sum_{\bs{X}^{1,\bs{w}'}\in\mathcal{X}}\Pr[\bs{X}^{1,\bs{w}'}|\bs{X}^1,\bs{w}'] V(\bs{X}^{1,\bs{w}'}) - \sum_{\bs{X}^{2,\bs{w}}\in\mathcal{X}}\Pr[\bs{X}^{2,\bs{w}}|\bs{X}^2,\bs{w}] V(\bs{X}^{2,\bs{w}}) + \sum_{\bs{X}^{2,\bs{w}'}\in\mathcal{X}}\Pr[\bs{X}^{2,\bs{w}'}|\bs{X}^2,\bs{w}'] V(\bs{X}^{2,\bs{w}'})
Since  $\bs{X}^1$ and $\bs{X}^2$ only differ in $A_{d,k}$ for $k\in\mathcal{K}^{\dagger}\triangleq\{k\in\mathcal{K}|\bs{w}_k=(1,2)\}$, by \eqref{eqn:trans_prob}, we can see that the next system states under control action $\bs{w}$ from $\bs{X}^1$ and $\bs{X}^2$ are the same. Thus, we have $A=C$.
For $B$ and $D$, if $j\in\mathcal{K}\setminus\mathcal{K}^{\dagger}$, by \eqref{eqn:trans_prob_k}, we can see that, $\bs{X}_{j,s}^{1,\bs{w}'}=\bs{X}_{j,s}^{2,\bs{w}'}$, $\bs{X}_{j,f}^{1,\bs{w}'}=\bs{X}_{j,f}^{2,\bs{w}'}$, and $\bs{X}_{j,un}^{1,\bs{w}'}=\bs{X}_{j,un}^{2,\bs{w}'}$ hold for all $\bs{w}'$.
If $j\in\mathcal{K}^{\dagger}$, we need to consider the following two cases under different $\bs{w}_j'$.
If $\bs{w}_j'=(0,0)$, then, by \eqref{eqn:trans_prob_k}, we can see that,
\begin{align*}
&\bs{X}_{j,un}^{1,\bs{w}'}=(\min\{A_{d,k}^1+1,\hat{A}_{d,k}\},\min\{A_{r,k}^1+1,\hat{A}_{r,k}\}, D_k^1),\nonumber\\
&\bs{X}_{j,un}^{2,\bs{w}'}=(\min\{\textcolor{black}{A_{d,k}^2+1},\hat{A}_{d,k}\},\min\{A_{r,k}^2+1,\hat{A}_{r,k}\}, D_k^2)
\end{align*}
% $\bs{X}_{j,un}^{1,\bs{w}'}=(\min\{A_{d,k}^1+1,\hat{A}_{d,k}\},\min\{A_{r,k}^1+1,\hat{A}_{r,k}\}, D_k^1)$ and $\bs{X}_{j,un}^{2,\bs{w}'}=(\min\{A_{d,k}^1+2,\hat{A}_{d,k}\},\min\{A_{r,k}^2+1,\hat{A}_{r,k}\}, D_k^2)$.
If $\bs{w}_j'=(1,1)$, then, we have
\begin{align}
&\bs{X}_{j,s}^{1,\bs{w}'}=\begin{cases}
(0,\min\{A_{d,k}^1+1,\hat{A}_{r,k}\}, L_k),  &\text{if}~D_k^1=1,\\
(\min\{A_{d,k}^1+1,\hat{A}_{d,k}\},\min\{A_{r,k}^1+1,\hat{A}_{r,k}\}, D_k^1-1),&\text{otherwise.}
      \end{cases}\\
&\bs{X}_{j,s}^{2,\bs{w}'}
=\begin{cases}
(0,\min\{A_{d,k}^2+1,\hat{A}_{r,k}\}, L_k),  &\text{if}~D_k^2=1,\\
(\min\{A_{d,k}^2+1,\hat{A}_{d,k}\},\min\{A_{r,k}^2+1,\hat{A}_{r,k}\}, D_k^2-1), &\text{otherwise.}
      \end{cases}\\
&\bs{X}_{k,f}^{1,\bs{w}'}= (\min\{A_{d,k}^1+1,\hat{A}_{d,k}\},\min\{A_{r,k}^1+1,\hat{A}_{r,k}\}, D_k^1),\\
&\bs{X}_{k,f}^{2,\bs{w}'}= (\min\{A_{d,k}^2+1,\hat{A}_{d,k}\},\min\{A_{r,k}^2+1,\hat{A}_{r,k}\}, D_k^2)
\end{align}
Thus, we can see that, $\bs{A}_d^{1,\bs{w}'}\succeq\bs{A}_d^{2,\bs{w}'}$, $\bs{A}_r^{1,\bs{w}'}\succeq\bs{A}_r^{2,\bs{w}'}$, and $\bs{D}^{1,\bs{w}'}=\bs{D}^{2,\bs{w}'}$, which imply $B\geq D$ according to Lemma~\ref{lemma:propertyV}.
Therefore, we can show that \eqref{eqn:J_property} holds.

Next, we prove Theorem~\ref{theorem:optimal} by using \eqref{eqn:J_property}.
Consider IoT device $k$, system action $\bs{w}=(\bs{w}_i)_{i\in\mathcal{K}}$ where $\bs{w}_k=(1,2)$, and system state $\bs{X}$ where $A_{d,k}=\phi_{\bs{w}}(\bs{X}_{-d,-k})$. Note that, we only to consider that $\phi_{\bs{w}}(\bs{X}_{-d,-k})< +\infty$. According to the definition of $\phi_{\bs{w}}(\bs{X}_{-d,-k})$, we can see that $J(\bs{X},\bs{w})\leq J(\bs{X},\bs{w}')$ holds for all $\bs{w}'\in\mathcal{W}$ and $\bs{w}'\neq \bs{w}$. Thus, we know that $\pi^*(\bs{X})=\bs{w}$. Now, consider another state $\bs{X}'$ where $A_{d,k}'\geq A_{d,k}$ and $\bs{X}_{-d,-k}'=\bs{X}_{-d,-k}$. To prove Theorem~\ref{theorem:optimal}, it is equivalent to show that $\pi^*(\bs{X})=\bs{w}$, i.e.,
\begin{equation}
J(\bs{X}',\bs{w})\leq J(\bs{X}',\bs{w}')
\end{equation}
holds for all $\bs{w}'\in\mathcal{W}$ and $\bs{w}'\neq \bs{w}$. By \eqref{eqn:J_property}, we can see that,
\begin{equation}
J(\bs{X}',\bs{w})- J(\bs{X}',\bs{w}') \leq J(\bs{X},\bs{w})- J(\bs{X},\bs{w}') \leq 0.
\end{equation}
Therefore, we obtain that $\pi^*(\bs{X}')=\bs{w}$, which completes the proof of Theorem~\ref{theorem:optimal}.

% \section{Proof of Lemma~\ref{lemma:separable}}\label{app:seperable}

\section{Proof of Lemma~\ref{lemma:propertyVk}}\label{app:per-value-function}
We prove Lemma~\ref{lemma:propertyVk} following a similar approach to Lemma~\ref{lemma:propertyV}.
First, we introduce the RVIA for the Bellman equation in \eqref{eqn:per-bellman}.
Denote $\hat{V}_{k}^n(\bs{X}_k)$ as the per-device value function at iteration $n$, where $n=1,2,\cdots$.
Then, we introduce the per-device state-action cost function under a randomized scheduling policy $\hat{\pi}_u$ at iteration $n$:
\begin{align}\label{eqn:J_kn_per}
\hat{J}_{k}^n(\bs{X}_k,v_k)&= A_{r,k}
+ \min_{v_k} \sum_{\bs{X}_k'\in\mathcal{X}_k}\mathbb{E}^{\hat{\pi}_{u}}\left[\Pr[\bs{X}'_k|\bs{X}_k,\bs{w}_k]\right] \hat{V}_{k}^n(\bs{X}'_k)
\end{align}
For each $(\bs{X}_k,v_k)$, the RVIA calculates $\hat{V}_{k}^{n+1}(\bs{X}_k)$ by:
\begin{equation}\label{eqn:RVIA_per}
  \hat{V}_{k}^{n+1}(\bs{X}_k)=\min_{v_k} \hat{J}_{k}^{n+1}(\bs{X}_k,v_k)- \min_{v_k}J_{k}^{n+1}(\bs{X}_k^{\dag},v_k),
\end{equation}
where $\bs{X}_k^{\dag}$ is some fixed state. Similar to \eqref{eqn:converge}, we also have
 \begin{equation}
   \lim_{n\to\infty}\hat{V}_{k}^n(\bs{X}_k)=\hat{V}_{k}(\bs{X}_k),~\forall \bs{X}\in\mathcal{X},\label{eqn:converge_per}
 \end{equation}
where $\hat{V}_{k}(\bs{X}_k)$ satisfies the Bellman equation in \eqref{eqn:per-bellman}.
% Let $\hat{\pi^*_{v_k}(\bs{X}_k)}$ be the control that attains the minimum of $\hat{J}_{k}^{n+1}(\bs{X}_k,v_k)$.

Now, consider two per-device state $\bs{X}_k^1$ and $\bs{X}_k^2$. To prove Lemma~\ref{lemma:propertyVk}, it is equivalent to show that, for any $\bs{X}_k^1, \bs{X}_k^2\in\mathcal{X}_k$ such that, $A_{d,k}^2 \geq A_{d,k}^1$, $A_{r,k}^2 \geq A_{r,k}^1$, and $D_k^2=D_k^1$,
\begin{align}\label{eqn:vkn}
\hat{V}_{k}^{n}(\bs{X}_k^2) \geq \hat{V}_{k}^{n}(\bs{X}_k^1),
\end{align}
holds for all $n=1,2,\cdots$.
This can be proved along the lines of the proof of Lemma~\ref{lemma:propertyV}. Therefore, by \eqref{eqn:converge_per}, we complete the proof of Lemma~\ref{lemma:propertyVk}.

\section{Proof of Theorem~\ref{theorem:suboptimal}}\label{app:suboptimal}

Based on Lemma~\ref{lemma:propertyVk}, by following the proof for \eqref{eqn:J_property} in Appendix~\ref{app:optimal}, we can easily show that, for any $\bs{X}^1,\bs{X}^2\in\mathcal{X}$, $\bs{w},\bs{w}'\in\mathcal{W}$ such that $\bs{A}_r^1=\bs{A}_r^2$, $\bs{D}^1=\bs{D}^2$, and
\begin{equation}
\begin{cases}A_{d,k}^1\geq A_{d,k}^2,  & \text{if}~\bs{w}_k=(1,2), \\
            A_{d,k}^1=A_{d,k}^2,  &\text{otherwise},
  \end{cases},
\end{equation}
for all $k\in\mathcal{K}$,
\begin{align}\label{eqn:J_property_sub}
\hat{J}(\bs{X}^1,\bs{w}) - \hat{J}(\bs{X}^1,\bs{w}') \leq \hat{J}(\bs{X}^2,\bs{w}) - \hat{J}(\bs{X}^2,\bs{w}').
\end{align}
holds. Then, consider IoT device $k$, system action $\bs{w}=(\bs{w}_i)_{i\in\mathcal{K}}$ where $\bs{w}_k=(1,2)$, system state $\bs{X}$ where $A_{d,k}=\hat{\phi}_{\bs{w}}(\bs{X}_{-d,-k})$, system state $\bs{X}'$ where $A_{d,k}'\geq A_{d,k}$ and $\bs{X}_{-d,-k}'=\bs{X}_{-d,-k}$.
By following the proof of Theorem~\ref{theorem:optimal} and by using \eqref{eqn:J_property_sub}, we can show that, $\hat{\pi}^*(\bs{X}')=\hat{\pi}^*(\bs{X})=\bs{w}$. We complete the proof of Theorem~\ref{theorem:suboptimal}.

\bibliographystyle{IEEEtran}
\bibliography{IEEEabrv,TWC_AoI}

\end{document}